\documentclass[aps,prc,twocolumn,nofootinbib,floatfix,groupedaddress,showpacs,amssymb,amsmath]{revtex4}
\usepackage{graphicx}
\usepackage{color}

\newcommand*{\AMeV}{\,A\,MeV}

\begin{document}



\title{Particle and light fragment
       emission in peripheral heavy ion collisions at Fermi energies}


%
%
\author{S.~Piantelli},
\author{P.R.~Maurenzig},
\author{A.~Olmi},
\thanks{corresponding author}
\email[e-mail:]{olmi@fi.infn.it}
\author{L.~Bardelli},
\author{A.~Bartoli},
\author{M.~Bini},
\author{G.~Casini},
\author{C.~Coppi},
 \author{A.~Mangiarotti},
\author{G.~Pasquali},
\author{G.~Poggi},
\author{A.A.~Stefanini},
\author{N.~Taccetti}
\author{E. Vanzi}

\affiliation{Sezione INFN and Universit\`a di Firenze, 
             Via G. Sansone 1, I-50019 Sesto Fiorentino, Italy}

\date{\today}

\begin{abstract}
A systematic investigation of the average multiplicities of light charged
particles and intermediate mass fragments emitted in peripheral and
semiperipheral collisions is presented as a function of the beam energy,
violence of the collision and mass of the system.
The data have been collected with the \textsc{Fiasco} setup in the
reactions $^{93}$Nb+$^{93}$Nb at 17, 23, 30, 38\AMeV\ and
$^{116}$Sn+$^{116}$Sn at 30, 38\AMeV.
The midvelocity emission has been separated from the emission of the 
projectile-like fragment.
This last component appears to be compatible with an evaporation
from an equilibrated source at normal density, as described by the 
statistical code \textsc{Gemini} at the appropriate excitation energy.
On the contrary, the midvelocity emission presents remarkable differences
for what concerns both the dependence of the multiplicities on the energy 
deposited in the midvelocity region and the isotopic composition of the
emitted light charged particles.
\end{abstract}


\maketitle

\section{Introduction \label{sec:intro}}

In recent years many works (see, e.g., 
\cite{Lukasik97,Plagnol99,Piantelli02,Milazzo02,Mangiarotti04} 
and references therein)
were devoted to the investigation of Light Charge Particle (LCP) and
Intermediate Mass Fragment (IMF) emission in semiperipheral and peripheral
heavy ion collisions at Fermi energies. 
It is well established that these reactions show mainly a binary
character, with two heavy remnants in the exit channel, possibly 
undergoing a subsequent sequential fission \cite{Stefanini95}.  
Among the various features of the emission, one in particular raised a 
lot of interest: a large amount of emission is located at midvelocity 
(i.e., close to the center-of-mass velocity), mainly for the IMFs but also 
for lighter particles
\cite{Plagnol99,Lukasik97,Bowman93,Montoya94,Toke96,Dempsey96,Piantelli02}. 
The origin of this phenomenon is still actively debated. 
Possible interpretations range from, e.g., a kind of  
neck rupture \cite{Lukasik97,Baran04}, due to mechanical and/or chemical 
instabilities, to a fast statistical emission from one of the heavy fragments
perturbed by the proximity of the other heavy remnant 
\cite{Botvina99,Botvina01}.

In a previous work \cite{Piantelli02}, by means of a three-body Coulomb
trajectory calculations, we put into evidence two components in the
experimental IMF emission pattern: a fast emission from the phase space 
region in between the two heavy remnants, somewhat reminiscent of a 
neck fragmentation or a participant-spectator model, and a later 
emission from the (possibly deformed) surface of the heavy fragments. 
The latter mechanism may be interpreted as the evolution of the fast
oriented fission mechanism \cite{Stefanini95,Casini93,Bocage00} toward very
large mass asymmetries. 
Similar conclusions were drawn in a more recent 
paper \cite{Defilippo05neck}, using 
the same basic approach of Coulomb trajectory calculations.

The midvelocity emission is particularly evident in very peripheral
reactions (b/b$_{\mathrm{graz}} \geq$ 0.9), where the kinematics of the 
collision is relatively simple and the multiplicity of LCPs (and IMFs) 
emitted from the two fully accelerated main fragments is quite small 
(about 0.3 LCP per event 
in Ref. \cite{Piantelli02}).
As a consequence, the investigation of
peripheral collisions is a fundamental tool in order to shed more light
on the production mechanism and on its evolution with the beam energy. 
The short interaction times typical of these collisions can also be 
used to investigate isospin diffusion and equilibration in collisions 
between nuclei with different N/Z ratio 
(see \cite{WCI06} and references therein).

In \cite{Mangiarotti04} we found that in peripheral collisions the energy 
stored inside the midvelocity matter and the excitation energy of the 
quasi-projectile have similar values.
This evidence suggested a larger energy density ($>$ 7 MeV/nucleon) 
in the midvelocity ``source'', with respect to the excited quasi-projectile
($\leq$ 2 MeV/nucleon),
since the mass localized at midvelocity is certainly smaller
(the value would even rise to $\approx$13 MeV/nucleon
if the mass of the midvelocity ``source'' 
were identified with the total mass of its emitted particles).
Similar conclusions were drawn from the comparison
of transverse velocities too \cite{Hudan05}.
If one assumes, as stated for example in \cite{Durand98},
that in central collision the multifragmentation
process starts at excitation energies of the source greater than
$\approx$3\AMeV, then the midvelocity emissions might be interpreted as
a first appearance of the multifragmentation process \cite{Baran04}.
Indeed it was also claimed that midvelocity fragments associated with 
midperipheral and central collisions present similar 
characteristics \cite{Hudan05}.

This paper presents a systematic investigation of the average 
multiplicities of LCPs and IMFs emitted in peripheral and semiperipheral 
collisions of symmetric systems at Fermi energies.
The evolution of the multiplicities is studied as a function of the 
excitation energy of the emitting ``source'', mass of the system and 
beam energy.
Section \ref{sec:setup} briefly describes the experimental setup 
\textsc{Fiasco} \cite{Bini03} and the investigated reactions.
Section \ref{sec:analysis} discusses the 
event selection and describes the analysis methods for 
separating the evaporative and midvelocity components of the emissions.
Section \ref{sec:results} presents the obtained results on the particle 
multiplicities and their scaling with the excitation energy and mass 
of the ``source'' and the beam energy, while conclusions are drawn in 
Section \ref{sec:concl}.
Some more technical points are presented in the Appendices.


\section{Experimental setup \label{sec:setup}}

The results presented in this paper were obtained in a systematic study 
\cite{Piantelli01,Piantelli02,Mangiarotti03,Mangiarotti04}
of heavy ion collisions performed at the Superconducting Cyclotron of
the Laboratori Nazionali del Sud of INFN in Catania.

Beams of $^{93}$Nb at  17.0, 23.0, 29.5 and 38.1\AMeV\  and 
of $^{116}$Sn at 29.6 and 38.1\AMeV\ were used to bombard 
$^{93}$Nb and $^{116}$Sn targets of about 200 $\mu$g/cm$^2$ thickness.
This paper concerns solely the symmetric collisions $^{93}$Nb + $^{93}$Nb 
and $^{116}$Sn + $^{116}$Sn, for which some parameters, calculated using 
the Bass interaction distance R$_{\mathrm{int}}$ \cite{Bass80}, are listed
in Table \ref{tab1}.

The experimental data have been measured with the \textsc{Fiasco} setup
(described with more details in \cite{Bini03,Piantelli01}), a multidetector 
particularly well suited for the study of peripheral and semiperipheral
reactions, where only few heavy remnants are produced.  
In fact, as a characteristic feature, this setup includes
24 large-area position-sensitive Parallel Plate Avalanche Detectors (PPAD)
\cite{Charity91,Stefanini95} covering about 70\% of the forward hemisphere. 
They are fully efficient for heavy fragments with Z$\agt$10
and are used to measure 
velocity vectors with very low detection thresholds 
($\sim$ 0.1\AMeV) and good accuracy 
(position and time-of-flight resolutions of 2-4 mm and 700 ps (FWHM), 
respectively, with flight-paths of about 3.5 m for $\theta \alt 10^{\circ}$). 
In this way it is possible to simultaneously detect both the
projectile-like fragment (PLF) and the very slow
target-like fragment (TLF) even in the most peripheral collisions and to 
perform a kinematic reconstruction of the events \cite{Casini89}.

Behind the most forward PPADs, a mosaic of 96 Silicon telescopes 
allows to measure the energy, the charge, and the final mass 
(by means of the time-of-flight) of the PLF.
Each telescope 
(with 28$\times$28 cm$^2$ active area)
consists of a 200$\mu$m $\Delta$E detector 
followed by a 500$\mu$m E$_{\mathrm{res}}$ detector.

Finally, the setup is completed by 182 three-layer phoswich scintillation 
telescopes, mounted behind most of the PPADs and covering about $30\%$ of 
the forward hemisphere (plus a reduced sampling in the backward hemisphere), 
which are devoted to the detection of light charged particles and
intermediate mass fragments. 
The phoswiches allow isotopic identification of Hydrogen (and 
in some cases of Helium) 
isotopes and charge identification of heavier products up to $Z\sim$15--20 
(with a threshold of about 3-3.5\AMeV) and give a direct measurement of the 
time of flight (and hence of the velocity) of all these reaction products, 
without any need for tricky and time-consuming energy calibrations of the
scintillators. 

   The experiment was performed together with the hodoscope 
   {\sc Hodo-Ct} \cite{Racitihodo} of the {\sc Temperature} 
   experiment \cite{Sfienti04}, but its data have not been used here.

The results presented in this paper are focused on binary events 
-- by far prevailing in peripheral and semiperipheral collisions -- 
with only two large reaction remnants (Z$\agt$10)
detected by the gas counters,
and on the associated multiplicities of LCPs and light IMFs with Z=3--7.

\begin{table}
 \caption{\label{tab1}  Some calculated reaction parameters}
 \begin{ruledtabular}
 \begin{tabular}{lrrrrrr}
    &    \multicolumn{4}{c}{$^{93}$Nb+$^{93}$Nb}   &
                              \multicolumn{2}{c}{$^{116}$Sn+$^{116}$Sn}\\ 
                                                \cline{2-5}  \cline{6-7}
     E/A   (MeV)           & 17   &  23  &  30  &  38  &  30  &  38  \\
                                                \cline{2-5}  \cline{6-7}
                           &      &      &      &      &      &      \\

E$_\mathrm{\,in}^\mathrm{\;c.m.}$
                     (MeV) &  791 & 1069 & 1374 & 1772 & 1715 & 2210 \\
$v_{\mathrm{rel}}$  (mm/ns)& 57.3 & 66.6 & 75.6 & 85.8 & 75.6 & 85.8 \\
$\theta_{\mathrm{graz}}^{\mathrm{Lab}}$
                     (deg) &  7.8 &  5.5 &  4.2 &  3.2 &  4.8 &  3.6 \\ 
$b_{\mathrm{graz}}$ (fm)
                           & 11.2 & 11.6 & 11.9 & 12.1 & 12.5 & 12.8 \\ 
$\ell_{\mathrm{graz}}$ 
                ($\hbar$)  &  470 &  568 &  660 &  762 &  865 & 1002 \\
$\sigma_{\mathrm{reac}}^{\mathrm{\,calc}}$ 
                 (mb)      & 3938 & 4260 & 4462 & 4621 & 4942 & 5143 \\ 

\end{tabular}
\end{ruledtabular}
\end{table}


 \section{Experimental data \label{sec:analysis}}   

\subsection{The choice of an ``ordering parameter'' \label{subs-orderparam}}

For a meaningful sorting of the data in homogeneous bins of increasing
centrality, one needs an experimental variable which is expected
to have a monotonic and possibly
narrow relationship with the impact parameter of the collision. 
This ``sorting'' variable should also be -as much as possible- 
independent of the studied observables, in order to avoid (or reduce) 
autocorrelation effects.

Global variables (like multiplicities, flow angles, transverse energies, 
etc.) are not used in the present work, as they are best suited
for experiments covering very large solid angles (close to 4$\pi$).
Therefore other variables, related to more particular aspects of the
reaction, must be considered.
Since the main subject of this paper is the emission of LCPs and IMFs in
binary collisions, it is natural to restrict the choice to variables which
make use of experimental information concerning the two main
reaction partners (like, e.g., the secondary charge $Z_{\mathrm{sec}}$ 
of the PLF residue, or its secondary velocity $v_{\,\mathrm{PLF}}$, or the 
relative velocity $v_{\mathrm{rel}}$ between PLF and TLF). 

Our choice is a variable \cite{Schroeder77} defined as 
\begin{equation}
  \mathrm{TKEL}  = \mathrm{E}_\mathrm{\,in}^\mathrm{\;c.m.}\, -\,
   \tfrac{1}{2}  \,  \Tilde{\mu}\,
   { v_{\mathrm{rel}}^{\,2}},      
  \label{eq:TKEL}
\end{equation}
where E$_\mathrm{\,in}^\mathrm{\;c.m.}$ is the center-of-mass (c.m.)
energy of the collision in the entrance channel,  
$v_{\mathrm{rel}}$ the reconstructed relative velocity and 
$\Tilde{\mu}$ the reduced mass calculated with the masses obtained from
the kinematic analysis.  

It is to be noted that -by definition- the kinematic method constrains
the primary masses of the two reaction partners to add up to the total mass
of the system.
Thus, while at low incident energies, where reactions are strictly binary,
TKEL may truly represent the ``Total Kinetic Energy Loss'' of the collision
 --~namely the total amount of kinetic energy transferred from the relative
 motion into internal energy of the colliding system~-- it is important to 
note that at Fermi energies, where the presence of
a sizable midvelocity emission causes an overestimation of the
kinematically determined $\Tilde{\mu}$, this interpretation is no longer
correct. 
Therefore, in this work 
{\em TKEL is used just as an ordering parameter for sorting the events
  in bins of increasing centrality}.

\begin{table}[b]
 \caption{\label{tab2}   Reaction cross sections}
 \begin{ruledtabular}
  \begin{tabular}{lrrrrrr}
    &    \multicolumn{4}{c}{$^{93}$Nb + $^{93}$Nb}   &
                              \multicolumn{2}{c}{$^{116}$Sn+$^{116}$Sn}\\ 
                                                \cline{2-5}  \cline{6-7}
 E/A   (MeV)    
    &     17    &     23    &     30    &     38    &     30    &     38 \\
                                                \cline{2-5}  \cline{6-7}
    &           &           &           &           &           &        \\
$\sigma_{2b}^{\mathrm{\,exp}}$ (mb)    
    &   3210    &   2840    &   2750    &   2810    &   2670    &   3010 \\
$\sigma_{3b}^{\mathrm{\,exp}}$ (mb) 
    &    640    &    940    &    940    &    790    &   1030    &    980 \\ 
$\sigma_{\mathrm{2b}}^{\mathrm{\,exp}} /
 \sigma_{\mathrm{reac}}^{\mathrm{\,calc}}$ 
    &    82\%   &    67\%   &   62\%    &    61\%   &   54\%    &   58\% \\
$\sigma_{\mathrm{2b+3b}}^{\mathrm{\,exp}} /
 \sigma_{\mathrm{reac}}^{\mathrm{\,calc}}$
    &    98\%   &    89\%   &   83\%    &    78\%   &   75\%    &   77\% \\

  \end{tabular}
 \end{ruledtabular}
\end{table}

A short discussion about the relationship of TKEL with other possible
sorting variables is also presented in Appendix A, 
together with an estimation of the correspondence between TKEL and 
impact parameter obtained both by means of model calculations 
and experimentally via a direct integration of the reaction cross section. 
We anticipate here that, for a given colliding system, equal values of 
TKEL correspond to similar impact parameters regardless of the bombarding 
energy, at least as far as (semi-)peripheral collisions are concerned.
When the system is changed (in our case from Nb+Nb to Sn+Sn), a given value 
of TKEL indicates a somewhat
larger impact parameter for the heavier system, 
as expected from the larger nuclear radii.


\subsection{The cross sections \label{subs-cross_sect}}

For each reaction, the simultaneous measurement (via a dedicated and 
suitably down-scaled ``singles'' trigger) of elastically scattered 
projectiles hitting the most forward gas detectors has been used for 
determining the conversion factor {\em (``millibarn-per-count'')}.
After correcting for the experimental filter, 
this allows to perform a quantitative estimate of the experimental 
cross sections pertaining to the detected two- and three-body events.
They are summarized, for all investigated reactions,
in the first and second row of Table \ref{tab2}, 
while the third row indicates the percentage of the calculated total 
reaction cross section which is accounted for by these two channels
(more details are given in Appendix A).

The third row shows that the exit channel with two heavy remnants remains
the dominant one even at the highest investigated energies, where it still 
accounts for more than half of $\sigma_{\mathrm{reac}}^{\mathrm{\,calc}}$,
thus demonstrating the persistence of the binary or quasi-binary character 
of the reactions.
The three-body channel, which is likely to be due to sequential fission 
or to the ``fast oriented fission'' \cite{Casini93,Stefanini95} of one of 
the two main fragments, adds an appreciable contribution.
For the $^{93}$Nb beams (where more beam energies have been studied)
it is interesting to note that at 17\AMeV\ the two- and three-body channels 
account for nearly 100\% (within errors) of the whole reaction cross section 
and that this percentage decreases with increasing bombarding energy.
This behavior may be due to the progressive opening of new reaction channels 
(four- or more-body reactions, possibly multifragmentation). 
However, even at the highest energy of 38\AMeV, the two- and three-body  
exit channels altogether still represent about 75\% 
of $\sigma_{\mathrm{reac}}^{\mathrm{\,calc}}$.

The main uncertainty on the quoted numbers concerns the cross sections 
for two-body events. 
It mainly arises from the difficulty of separating elastic 
and quasi-elastic scattering in the vicinity of the grazing angle.
The experimental data have been integrated starting from the calculated 
grazing angles of Table \ref{tab1}, after having verified that they agree
with the experimental ``quarter-point'' angles within a few tenths of degree.
Such an uncertainty in angle corresponds to an uncertainty of the order of 
200-300 mb on the cross sections and of about 5--6\% on the
percentages in the last row.


%
\subsection{The emission pattern \label{subs-expdata}}

 \begin{figure*}[t]
  \includegraphics[height=63mm,bb= 0 0 655 400,clip]{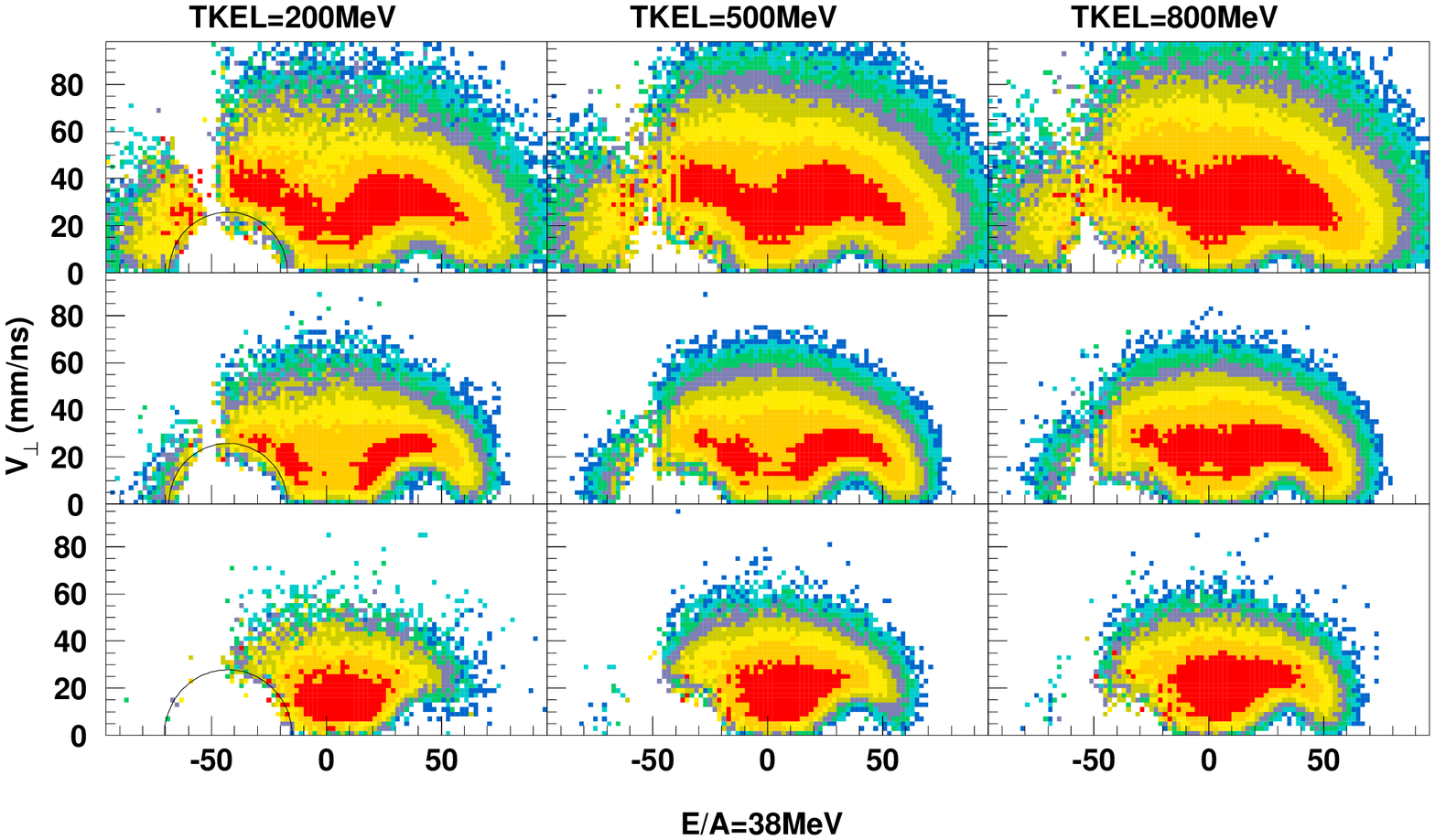}
  \includegraphics[height=63mm,bb=50 0 500 400,clip]{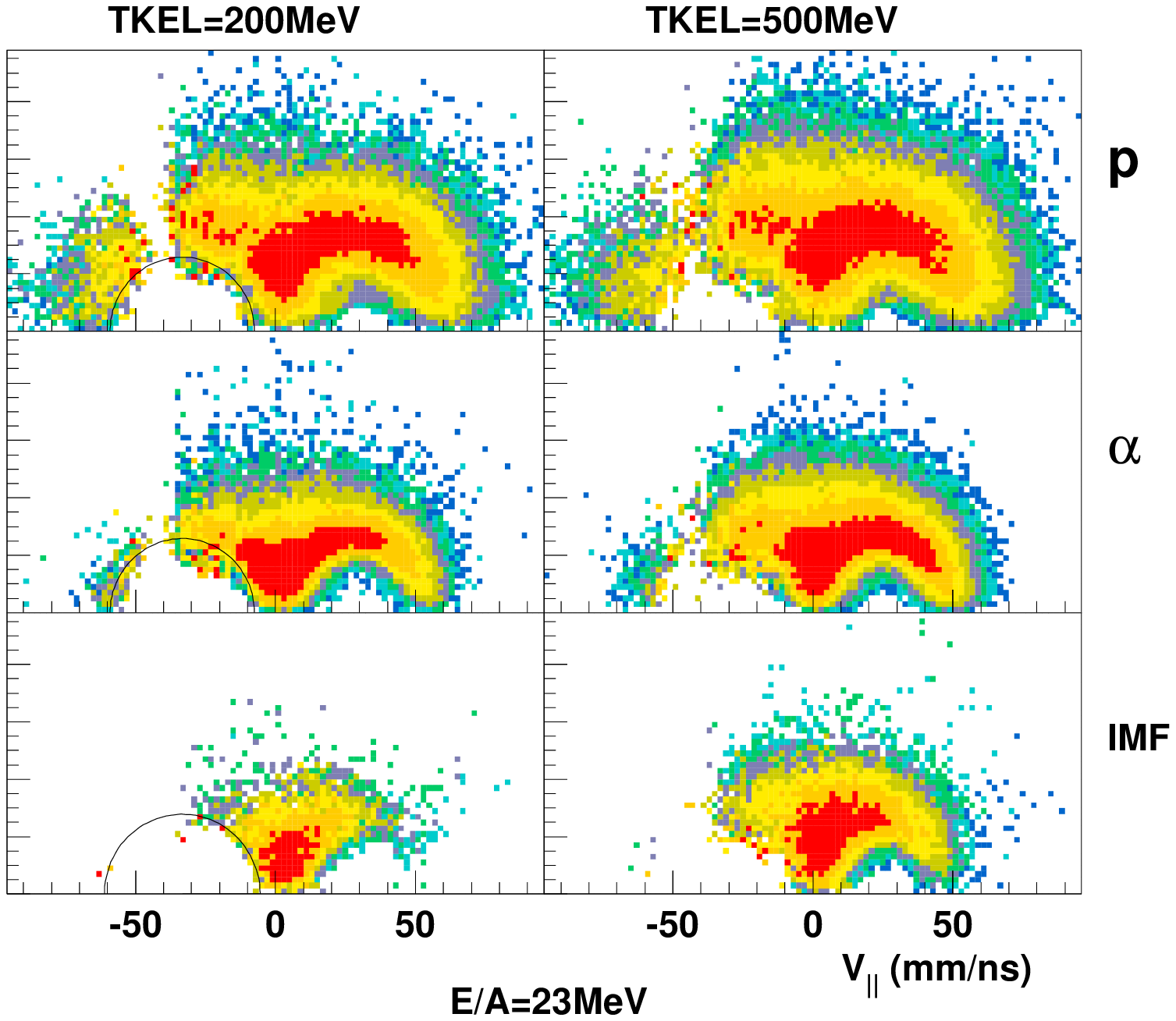}
  \caption
  {\label{fig:occhi}
  (color)
    Experimental yields (corrected for the setup efficiency) of protons
   (upper row), $\alpha$ particles (middle row) and IMFs 
    with $Z$=3--7 (lower row) for $^{93}$Nb + $^{93}$Nb at 
    38 (left panels) and 23\AMeV\ (right panels). 
    Yields -- with logarithmic level spacing in factor-of-2 steps -- 
    are in the plane ($v_{\,\parallel}, v_{\,\perp}$), with respect to the 
    PLF-TLF separation axis, the origin being in the center-of-mass system.
    Columns refer to 100-MeV wide TKEL bins centered, from left to right, 
    at 200, 500, 800 MeV (at 38\AMeV) and 200, 500 MeV (23\AMeV). 
    The circles in the TKEL=200\,MeV panels show regions affected by 
    velocity thresholds (due to the thin fast-plastic scintillators of 
    the phoswich telescopes, see \cite{Bini03}). 
  }
 \end{figure*}

The shape of the emission pattern of LCPs and IMFs can be best 
appreciated from the distribution of the experimental yields in the 
($v_{\,\parallel}$,$v_{\,\perp}$) plane, where $v_{\,\parallel}$ and 
$v_{\,\perp}$ are the velocity components (in the c.m. reference system)
parallel and perpendicular, respectively, to the asymptotic PLF-TLF
separation axis (for the TKEL range addressed in this work, the separation
axis remains in most cases within about 10$^\circ$ from the beam axis).
As the solid angle coverage of the \textsc{Fiasco} setup, although large, 
is however significantly smaller than 4$\pi$, it is first necessary to 
correct the data for the limited geometrical coverage
of the setup and for the low-energy
identification thresholds \cite{Piantelli01}.  

The correction \cite{Piantelli01,Mangiarotti03} is obtained from a 
Monte Carlo simulation which produces - for each particle species - 
a random isotropic emission.
In fact, it has been verified that, thanks to the large acceptance and 
axial symmetry of the setup, the obtained correction is largely independent
of the specific emission pattern used in the simulation. 
Therefore the emission adopted in the simulation homogeneously fills a 
sphere in phase space,
centered in the c.m. origin with radius $v_{\mathrm{max}}$= 120 mm/ns, 
thus covering all regions which are relevant for the processes under study.
The velocity vector of each LCP or IMF is then transformed into
the laboratory reference frame and the appropriate experimental filter of 
the \textsc{Fiasco} setup is applied
(geometry and identification thresholds of the phoswich detectors).
In order to mimic as closely as possible the binary reaction step, all its
relevant parameters (like the orientation of the PLF-TLF separation axis 
and the PLF and TLF c.m. velocities) are those obtained with the 
kinematic reconstruction directly from the measured two-body events. 
Using these parameters, the procedure allows to determine, for each cell 
in the ($v_{\,\parallel}$, $v_{\,\perp}$, $\phi$) space and for successive 
bins in TKEL, an average correction factor, which is obtained as the ratio 
between the number of generated particles and the number of particles 
surviving the experimental filter of the setup.
(Here $\phi$ is the out-of-plane angle with respect to the reaction plane;
this coordinate was explicitly considered for a better efficiency
correction in case of possible out-of-plane anisotropies in the
experimental data, e.g., due to angular momentum effects).
When analyzing the actual data, the experimental yields of each particle 
species are multiplied, cell-by-cell, by the now described correction factors.

As an example of the obtained distributions, Fig.~\ref{fig:occhi} shows 
the efficiency-corrected experimental yields of protons, 
$\alpha$ particles\footnote{
     Among the Helium isotopes, $^4$He is the dominant one, while $^3$He 
     accounts for a few percent of the total (it is visible on the left 
     tail of $^4$He only in the phoswiches with the best resolution) 
     and $^6$He cannot be singled out. Therefore, although the data 
     presented in this paper refer to all He isotopes, they are 
     representative of the behavior of $\alpha$ particles only.}
and IMFs with $Z$=3--7 (first, second and third row, respectively) in the 
reaction $^{93}$Nb + $^{93}$Nb at 38 and 23\AMeV\ for three and two bins 
of TKEL, respectively.
In this presentation, in absence of instrumental effects,
the average positions of the PLF and TLF emitters 
lie -by definition- on the horizontal axis, symmetrically with respect 
to the c.m. origin. 
One clearly sees characteristic circles around the positions of PLF and
TLF, indicating a Coulomb-dominated emission from these sources, as it is
expected for the sequential decay of such highly excited systems.
(In the case of TLF, the inner part of the Coulomb circles 
is marginally affected by the velocity thresholds.)
It has also to be noted that, since yields are presented in 
Fig.~\ref{fig:occhi} -- and not invariant cross sections --
the intensity of an isotropic emission must gradually decrease along 
the Coulomb circles while approaching the $v_{\,\parallel}$-axis, 
until it vanishes when it reaches this axis. 
At parallel velocities intermediate between those of PLF and TLF, one 
observes an additional contribution, the so-called ``midvelocity'' 
(or ``neck'') emission.
Although present for all particle species, this emission is 
particularly evident and important for $\alpha$ particles and 
even more for IMFs.

The \textsc{Fiasco} setup has a much better solid angle coverage in 
forward direction, where -in addition- thresholds effects do not play 
any practical role, because the energies of all particles in the 
lab-system are greater than $\approx$~4\AMeV\ already for the collision 
at 17\AMeV.
On the contrary, the solid angle coverage in backward
direction is limited. It has to be noted that
phase space cells with small average efficiency have large correction 
factors, which amplify the statistical fluctuations of the experimental 
data, and cells with zero efficiency cannot be corrected at all.
This is the reason why in the backward hemisphere of the laboratory
reference frame 
(corresponding in c.m. to $v_{\,\parallel}\alt -\,43$\,mm/ns for the 
reaction at 38\AMeV\ and to $v_{\,\parallel}\alt -\,33$\,mm/ns for
that at 23\AMeV) the applied corrections are not
as effective as in the forward hemisphere.

Finally, as the results presented in this paper concern
LCPs and IMFs emitted in peripheral {\em binary} reactions, 
the data need to be corrected for the presence of a background of events 
with a higher multiplicity of heavy fragments, of which only two
have been detected.
This background of incomplete events (a few percent)
has been estimated using the measured three-body events 
(with the procedure described in Ref.~\cite{Casini89}) 
and subtracted from all the data presented in this paper.


\subsection{The particle multiplicities \label{subs-multot}}

The average multiplicities of charged particles  
were obtained from the (efficiency corrected and background subtracted) 
experimental distributions of p, d, t, $\alpha$ and IMFs, some examples 
of which are shown in Fig. \ref{fig:occhi}.

An advantage of using symmetric collisions is that the forward-going
particles (those with $v_{\,\parallel}\geq 0$ in c.m.)
must have the same average characteristics as the backward-going ones.
Therefore, because of the already mentioned much better quality of the data,
all presented multiplicities refer only to particles emitted in the
forward hemisphere of the c.m. reference frame 
(of course, the average multiplicities for the 
{\em whole}
colliding system can be obtained by simply doubling the presented values). 

 \begin{figure}[t]
  \includegraphics[width=57mm,bb=20 20 350 410]{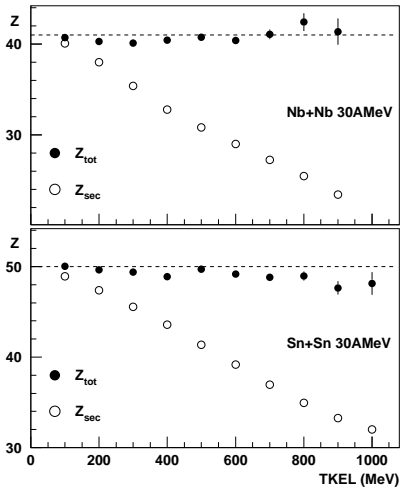}
  \caption
  {\label{fig:ztot}
   Average total charge Z$_{\mathrm{tot}}$ of forward-going products 
   (full dots) and average secondary charge Z$_{\mathrm{sec}}$ of PLF 
   (open dots) in $^{93}$Nb + $^{93}$Nb and $^{116}$Sn + $^{116}$Sn 
   at 30\AMeV.
   Dashed lines indicate the value of the projectile charge.
  }
 \end{figure}

 \begin{figure}[t]
  \includegraphics[width=80mm,bb=10 10 515 830,clip]{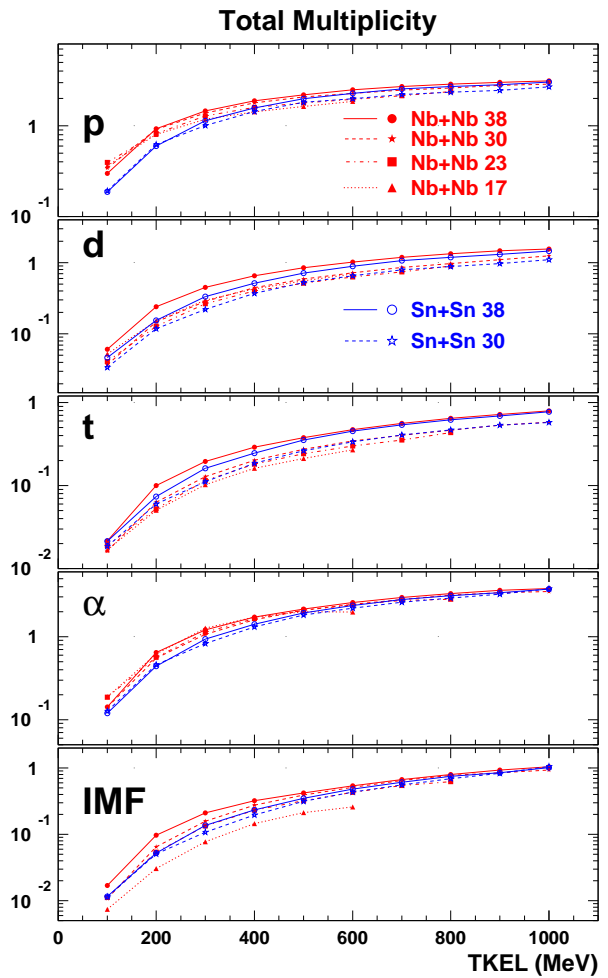}
  \caption
  {\label{fig:multot}
   (color online) 
   Experimental efficiency-corrected average multiplicities of 
   (forward-emitted) p, d, t, $\alpha$ and IMFs ($Z$=3--7) as a function 
   of TKEL, for $^{93}$Nb+$^{93}$Nb (full symbols) 
   and $^{116}$Sn+$^{116}$Sn (open symbols). 
   The bombarding energies are 17 (triangles, dotted lines),
   23 (squares, dot-dashed lines), 30 (stars, dashed lines) and
   38\AMeV\ (circles, full lines); the lines are to guide the eye.
  }
 \end{figure}

The symmetry of the system also allows to check the quality of the 
applied efficiency corrections.
In fact, whatever the reaction mechanism, by adding up the charges of
forward emitted LCPs and IMFs to the charge of the PLF residue, one 
should reproduce -on average- just the projectile charge (of course, 
this can be done only for that subset of the events in which the PLF 
residue was detected and identified in one of the Silicon telescopes). 
As an example of the quality of this check, the full dots of 
Fig.~\ref{fig:ztot} show the average total forward-emitted charge for the
$^{93}$Nb + $^{93}$Nb and $^{116}$Sn + $^{116}$Sn collisions at 30\AMeV: 
within about half a charge unit, the value of the projectile 
(dotted line) is well reproduced, thus giving confidence in the present 
analysis.
The rapidly widening gap between the full dots and the open ones 
(which represent the average secondary charge Z$_{\mathrm{sec}}$ of 
the detected PLF) gives a visual indication of the rising amount of 
charges removed by the LCP and IMF emissions with increasing TKEL.

The obtained 
average multiplicities (of forward-going particles) are shown in Fig. 
\ref{fig:multot} for the systems $^{93}$Nb + $^{93}$Nb (full symbols) and 
$^{116}$Sn + $^{116}$Sn (open symbols) at the bombarding energies of 17, 
23, 30 and 38\AMeV\ (triangles, squares, stars and circles, respectively).
The statistical errors are smaller than the symbols and the lines are 
just to guide the eye.

Although data corresponding to more violent collisions have been
acquired too, the present analysis is limited to peripheral and
semiperipheral collisions through a restriction of the TKEL range (from
$\leq$600 MeV for the reaction at 17\AMeV\ up to $\leq$1000 MeV for
those at 38\AMeV). 
This has the twofold advantage of allowing a clear and unambiguous
distinction between PLF and TLF (in c.m. the forward-flying heavy 
remnant is always the PLF) and of limiting the study to regions where
the binary exit channel is the dominant one. 
In the considered TKEL ranges the binary exit channel remains
approximately symmetric in mass, as indicated by comparable values 
of the c.m. velocities of the two main fragments.
Only in the last considered TKEL bins weak tails of more asymmetric 
mass splittings appear. 
In the analysis, the requirement 0.4 $\leq v^{\mathrm{c.m.}}_{\mathrm{PLF}}/
(v^{\mathrm{c.m.}}_{\mathrm{PLF}}+v^{\mathrm{c.m.}}_{\mathrm{TLF}})\leq\;$0.6 
has been used to reject these tails, which are strongly contaminated by 
incompletely detected three-body events.

Qualitatively, all multiplicities display a similar behavior, with a
strong dependence on TKEL and a much weaker one on bombarding energy and 
mass of the system.
Starting from small TKEL they all increase rapidly but then tend 
to flatten at large TKEL values.
Over the range of TKEL considered here, the multiplicities span about
one order of magnitude -or slightly more- for the most abundant species 
(protons and $\alpha$ particles, with multiplicities up to several
particles per event) and about two orders of magnitude for the rarer
reaction products (which barely reach multiplicities of one per event 
at the highest values of TKEL).


\subsection{Evaporative and midvelocity emissions \label{subs-statmidv}}

\begin{figure}[t]
  \includegraphics[width=80mm,bb=0 0 567 530,clip]{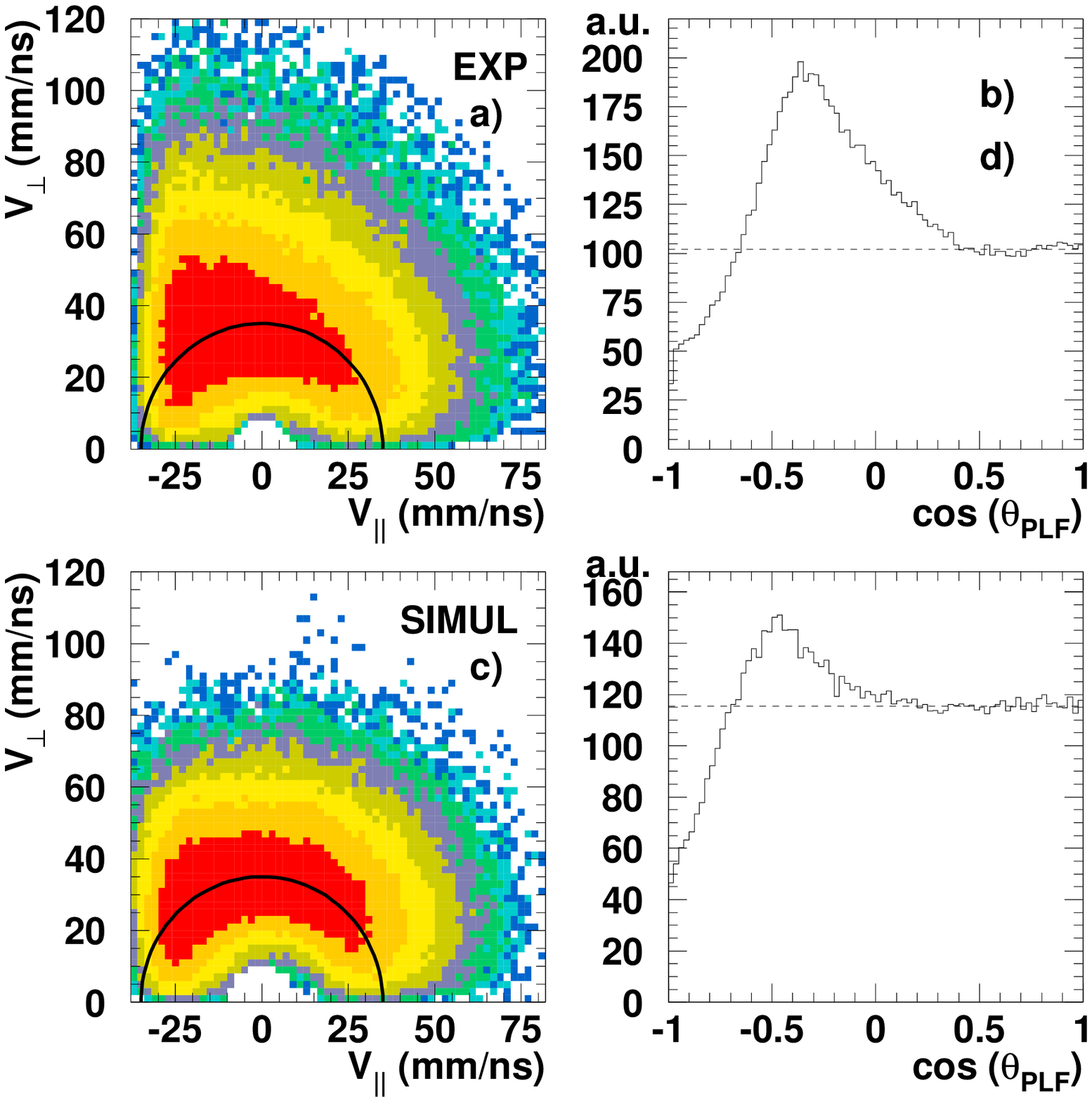}
  \includegraphics[width=50mm,bb=0 0 505 220,clip]{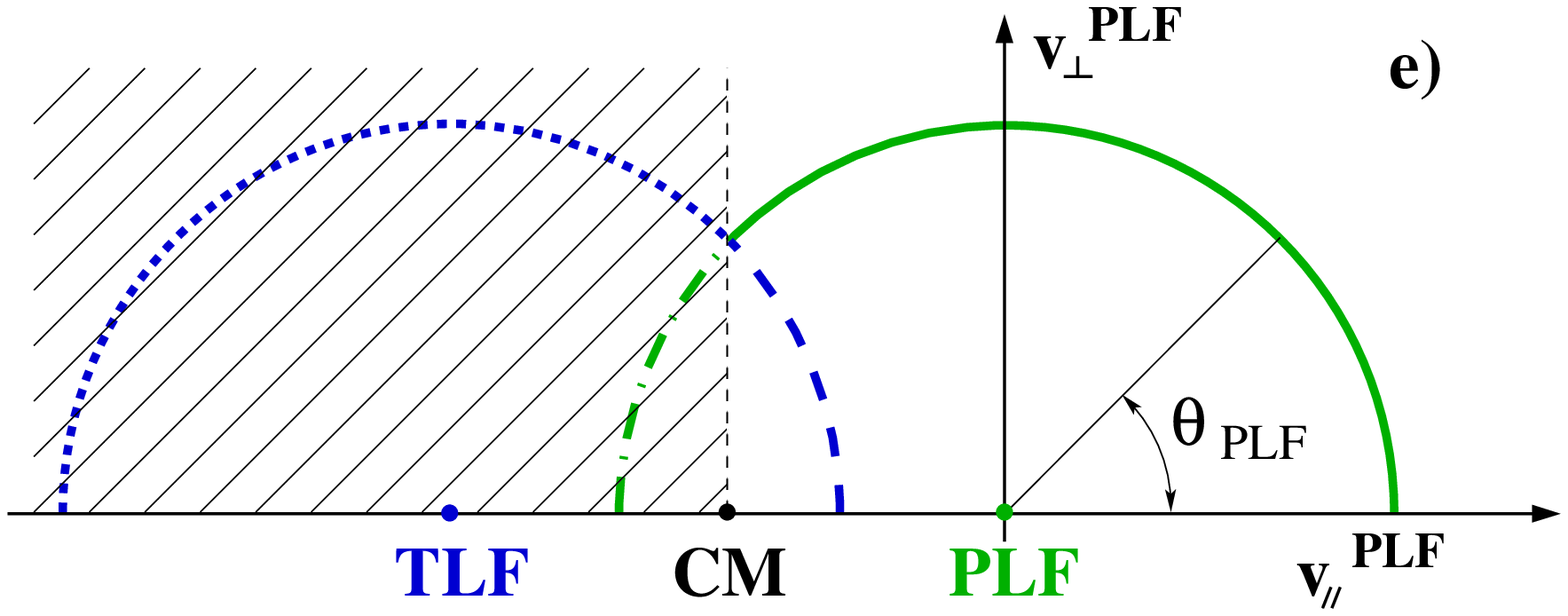}
  \caption
  {\label{fig:occhi-cost}
   (color) 
   Experimental (a) and simulated (c) yields of protons in the 
   ($v_{\,\parallel}^{\,\mathrm{PLF}}, v_{\,\perp}^{\,\mathrm{PLF}}$) 
   plane (with origin in the PLF reference frame) and 
   corresponding experimental (b) and simulated (d) distributions 
   of $d\sigma/d\cos(\theta_{\mathrm{PLF}})$
   for TKEL=800 MeV in the reaction $^{93}$Nb+$^{93}$Nb at 38\AMeV.
   Only forward-going particles (i.e., with $v_{\,\parallel}$ in the 
   lab-system larger than the c.m. velocity) are considered 
   --as sketched in (e)--  and this causes the yields to vanish 
   for $v_{\,\parallel} \alt$ -35 mm/ns in panels (a) and (c).
  }
 \end{figure}

In order to disentangle the midvelocity emission from the sequential
evaporation of the PLF, it is convenient to make a coordinate 
transformation into the reference frame of the PLF, namely a frame 
with the $v_{\,\parallel}$-axis still oriented along the asymptotic 
PLF-TLF separation axis, but with the origin on the PLF itself. 
A relativistic Lorentz transformation has been applied (instead of 
simpler Galilean one), in order to avoid distortions of the angular 
distribution of the fastest particles; this comes out to be necessary 
in particular for protons, which may have lab-velocities as large 
as 30\% of the speed of light.    
An example of the obtained yield in the 
($v_{\,\parallel}^{\,\mathrm{PLF}},v_{\,\perp}^{\,\mathrm{PLF}}$) plane 
is shown in Fig.~\ref{fig:occhi-cost}(a) for protons, at TKEL= 800 MeV, 
in the reaction $^{93}$Nb + $^{93}$Nb at 38\AMeV. 
The corresponding angular distribution, presented in 
Fig.~\ref{fig:occhi-cost}(b), is obtained by plotting the proton
yield as a function of $\cos(\theta_{\mathrm{PLF}})$, 
where $\theta_{\mathrm{PLF}}$ is the polar emission angle between the
velocity of the protons in the PLF frame and the PLF-TLF separation 
axis, as sketched in Fig.~\ref{fig:occhi-cost}(e):
thus $\cos\,(\theta_{\mathrm{PLF}}) = 1$ corresponds to a forward
emission along the flight direction of the PLF in the c.m. system.

For comparison, Fig.~\ref{fig:occhi-cost}(c) and (d) show the 
corresponding results obtained with a Monte Carlo simulation, where only 
an isotropic evaporation from the fully accelerated PLF and TLF is
considered, as sketched by the two circles in Fig.~\ref{fig:occhi-cost}(e).  
At small TKEL the emissions from PLF and TLF are well 
separated from each other and the 
$\cos\,(\theta_{\mathrm{PLF}})$-distribution is flat.
At larger TKEL, on the contrary, the distribution presents a dip at
$\cos\,(\theta_{\mathrm{PLF}})$= -1 and a bump 
at less negative values of $\cos\,(\theta_{\mathrm{PLF}})$,
as shown in Fig.~\ref{fig:occhi-cost}(d).
This distortion of the backward part of the angular distribution is due 
to the merging of the two evaporative emissions (from PLF and TLF) when 
the relative velocity of the two sources is reduced: as shown in 
Fig.~\ref{fig:occhi-cost}(e), selecting the forward-going particles 
(in c.m.) one misses the dash-dotted part of the PLF circle, 
but includes the dashed part of the TLF circle.
Nonetheless, thanks to the symmetry of the system, the yields and the 
multiplicities are correctly evaluated.
In fact, with respect to the flat behavior of the forward part of 
the $\cos\,(\theta_{\mathrm{PLF}})$-distribution 
[dashed line in Fig.~\ref{fig:occhi-cost}(d)],
the excess area in the bump neatly compensates the deficit around -1.

When comparing the experimental angular distribution of 
Fig.~\ref{fig:occhi-cost}(b) with the simulated one of 
Fig.~\ref{fig:occhi-cost}(d), it becomes evident that:
(i)
     The shapes are qualitatively similar.
(ii)
     The forward part of the experimental angular distribution, in the
     vicinity of $\cos\,(\theta_{\mathrm{PLF}})$ = 1 is indeed flat, 
     as expected for an evaporative (nearly isotropic) component.
(iii)
     The broad bump around $\cos(\theta_{\mathrm{PLF}})$ = - 0.5 
     overcompensates the dip near -1, thus confirming the presence of
     an additional backward emission (with respect to the PLF frame), 
     namely an emission in the midvelocity region of the phase-space.
(iv)
     The tail of the bump extends well into the forward hemisphere 
     in the PLF frame  \cite{Mangiarotti04}.

Thus, as already noted in \cite{Mangiarotti04}, the analysis of the 
experimental angular distribution suggests the superposition of two 
emission components, namely one from the PLF and one from ``midvelocity''.
As it will be shown in Sec. \ref{subs-sqrte}, there are good arguments for
interpreting the first component as an evaporation from the excited PLF; 
therefore the two terms, evaporation and PLF-emission, will be used 
interchangeably in the rest of this paper.
On the basis of the previous point (ii), a decomposition can be attempted,
provided that the shape of the whole evaporative component can be 
reliably estimated. 
Since evaporation must be forward-backward symmetric in the frame of the 
emitter, it is usual (see, e.g., \cite{Plagnol99,Dagostino99})
to estimate the total evaporation from the PLF by 
simply doubling its forward emission.
This may work at higher bombarding energies (where evaporation from 
PLF/TLF and midvelocity emissions are well separated in phase space),
but the above point (iv) shows that, even at the highest 
energy of 38\AMeV, the usual procedure would result in a significant
contamination from the midvelocity emission. 
Which part of the angular distribution can be considered sufficiently 
clean depends on the bombarding energy, on the chosen TKEL bin and also 
on the considered particle species.
However, for the reactions of this paper, 
the range $0^\circ \leq \theta_{\mathrm{PLF}} \leq 45^\circ$
(i.e., $0.7 \leq \cos(\theta_{\mathrm{PLF}}) \leq 1$) 
is a reasonable compromise which can be used in almost all cases.
(We verified that the total evaporative multiplicities remain stable 
within $\pm$10\% if the data are taken, e.g., in the narrower range 
$0^\circ \leq \theta_{\mathrm{PLF}} \leq 30^\circ$ while progressively
larger deviations appear if data beyond $45^\circ$ are included, see 
Appendix B.).

In order to estimate, from the data measured at forward angles, the 
total yield of the PLF component, it is necessary to make some hypothesis 
on the angular distribution $d\sigma/d\cos(\theta_{\mathrm{PLF}})$, 
which depends on the spin of the emitter.
In fact this distribution is flat only in case of an isotropic 
evaporation from a zero-spin source, while in case of non-zero spin the 
evaporation tends to concentrate in a plane perpendicular to the spin 
direction, giving origin to an U-shaped $\cos(\theta_{\mathrm{PLF}})$
distribution: the larger the spin 
(or the heavier the evaporated particle), the stronger the effect.
According to a detailed study of the correlations of the emitted
particles with both the PLF and TLF \cite{Mangiarotti03,Mangiarotti05}, 
it was estimated that in the reaction $^{93}$Nb + $^{93}$Nb 
at 38\AMeV\ the spin of the PLF is negligible at low TKEL, but 
rises to $15 \pm 5 \hbar$ and $30 \pm 10 \hbar$ at TKEL $\approx$ 500 and 
800 MeV, respectively.
However, the shape of the experimental distribution is
not very sensitive to the spin value.
In fact, as verified with Monte Carlo simulations, even in case of
an initial strong alignment of the spin perpendicular to the reaction 
plane, the anisotropy of the experimental angular distribution is reduced
-- with respect to the theoretical one \cite{Dossing81,Moretto81} --
by the misalignment of the spin (caused by the particle evaporation)
and by the fluctuations of the reconstructed reaction plane.
Both effects increase with increasing number of particles 
emitted along the evaporation chain and hence with increasing TKEL.
Therefore the yields for the evaporative component have been obtained 
by assuming a flat distribution (estimated from the data in the forward 
range $\cos(\theta_{\mathrm{PLF}}) \geq 0.7$) and applying a correction 
(estimated with Monte Carlo methods) which is at most of the order of 15\%.

 \begin{figure*}
  \begin{tabular}{c c}
    \includegraphics[width=80mm,bb=10 10 515 830,clip]{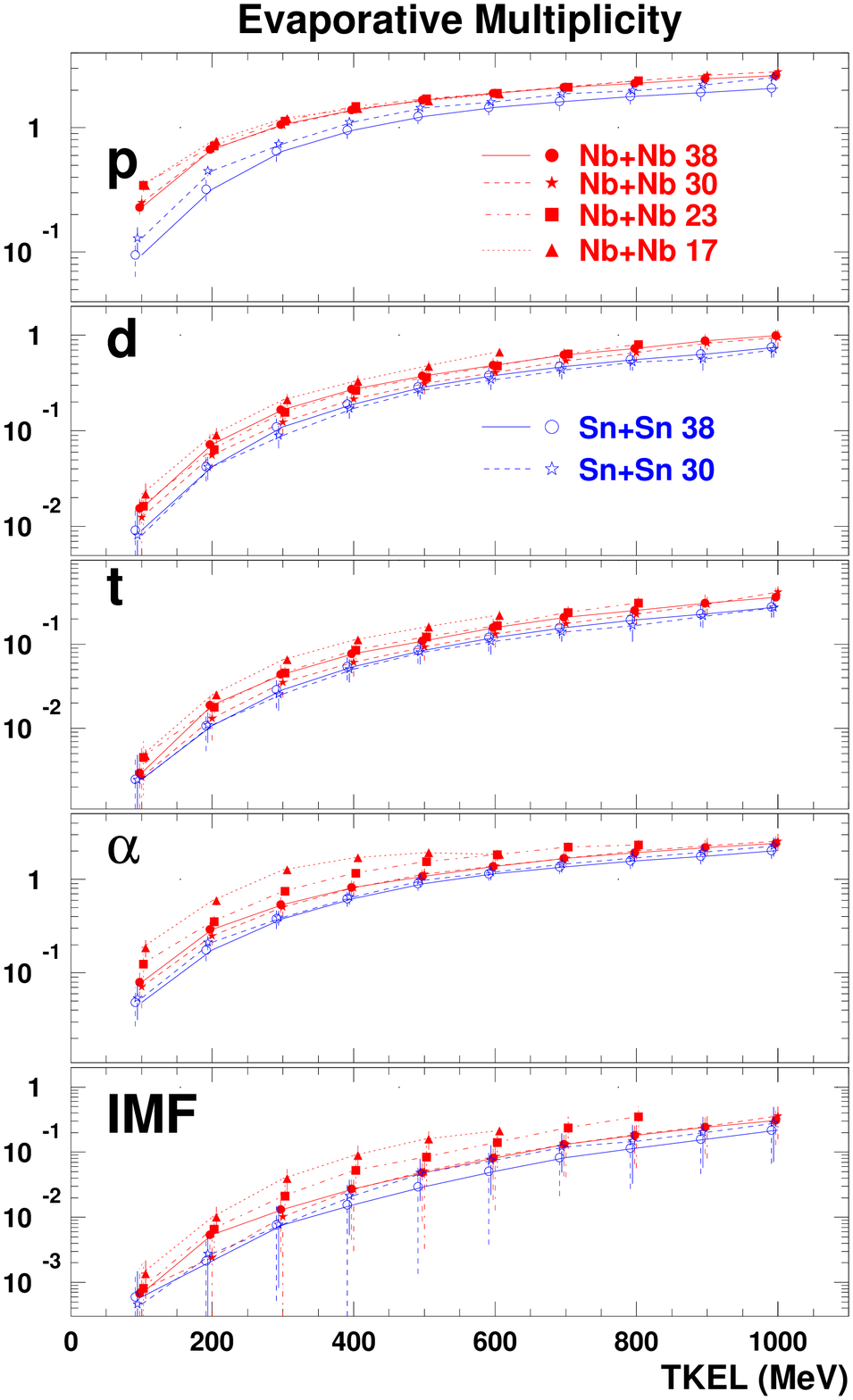}
   &\includegraphics[width=80mm,bb=10 10 515 830,clip]{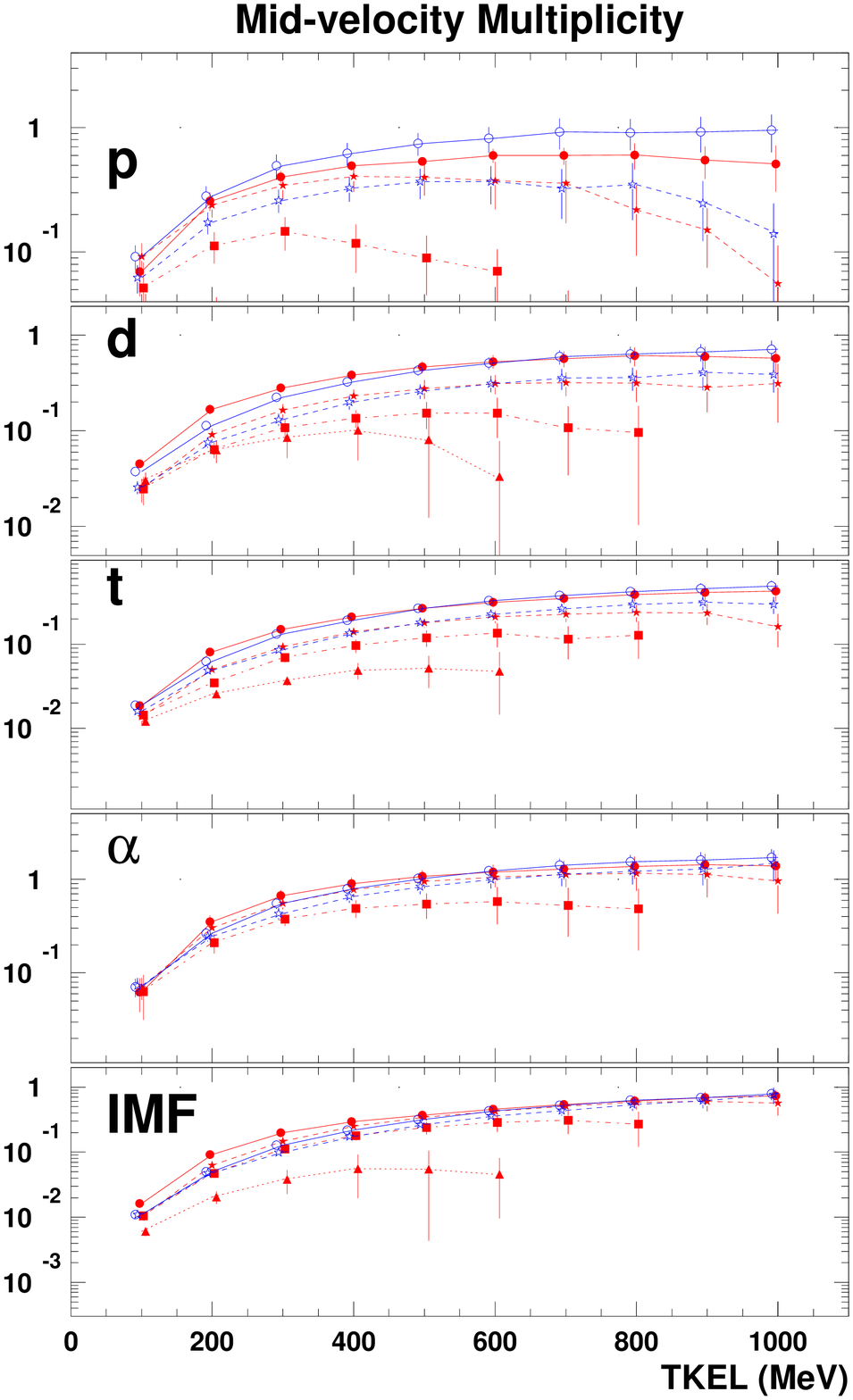}
  \end{tabular}
  \caption{
  \label{fig:moltepl} 
  (color online)
   Experimental, efficiency-corrected evaporative (left) and midvelocity
   (right) multiplicities of forward-emitted p, d, t, $\alpha$ and IMFs 
   ($Z$=3--7) as a function of TKEL, for $^{93}$Nb+$^{93}$Nb 
   (full symbols) and $^{116}$Sn+$^{116}$Sn (open symbols). 
   Error-bars indicate not statistical errors 
   (they are smaller than the symbols), but rather variations of 
   multiplicity values with analysis procedure (see text).
   The symbols, with the same meaning as in Fig. \ref{fig:multot},
   are slightly displaced horizontally in order to 
   appreciate the individual error bars at a given TKEL.
  }
 \end{figure*}

Once an estimate of the emission of particles from the PLF
has been obtained, the midvelocity component might 
be derived from the total multiplicities of Fig. \ref{fig:multot} using 
a subtraction procedure \cite{Plagnol99}.
However, for an unbiased determination of the evaporative component
one should take into account that, due to momentum conservation, the 
emission of particles -- especially of heavy ones -- from a source of 
finite mass produces recoil effects which perturb the velocity of 
the source itself.
Altogether, the net result of a chain of evaporation steps is a 
superposition of uncorrelated perturbations leading just to a smearing 
of the original source velocity, with an average null effect. 
However if the heavy remnant is observed in coincidence with a particle
emitted in a specific direction, this fact introduces a correlation 
between the velocities of the two objects.
In particular, if one selects events characterized by the emission 
of certain particles in a restricted angular range in the source frame, 
in those events the perturbation of the source velocity will have a 
non-zero average value.

The requirement that a given particle be emitted in forward direction
selects heavy residues which systematically recoil in backward direction.
This causes a systematic decrease of the c.m. velocity of the residue 
(and hence an overestimation of TKEL), while all the other unmeasured or 
unselected particles contribute only to the smearing of the data.
As a consequence, the PLF multiplicities evaluated in a given bin 
of TKEL do actually pertain to a range of somewhat smaller TKEL values.
Or, in other words, they erroneously include some contribution from events 
belonging to the previous bin (lower TKEL) and miss some contribution 
from events which are classified in the following one (larger TKEL).
It is worth noting that the heavy residue recoils in the (opposite) 
forward direction when it evaporates particles at angles 
$\theta_{\mathrm{PLF}} \geq 90^{\circ}$, 
which then merge with the midvelocity emissions.
In this case the c.m. velocity of the residue increases systematically 
and the resulting TKEL is underestimated. 
This has to be taken properly into account before subtracting the 
evaporative component from the total multiplicities. 


\section{Results \label{sec:results}}

\subsection{Particle multiplicities \label{subs-mult}}

The evaporative ($\mathcal{M}_{\mathrm{evap}}$) and 
midvelocity ($\mathcal{M}_{\mathrm{midv}}$) multiplicities 
of forward-going particles are presented, as a function of TKEL, 
in the left and right part of Fig.~\ref{fig:moltepl}, respectively. 
The multiplicities refer --with the same symbols-- to the same 
experimental data used for Fig.~\ref{fig:multot}.
The vertical scales differ for different particles, but they are the 
same for the evaporative and midvelocity components of a given particle.
It is also worth noting that at 17\AMeV\ it is not possible to reliably
extract the midvelocity multiplicities $\mathcal{M}_{\mathrm{midv}}$ 
for protons and $\alpha$ particles, because of the overwhelming 
contribution of the evaporative part.

The error bars of Fig. \ref{fig:moltepl} do not represent statistical 
errors (which are usually smaller than the symbols).
Instead, at each point, they show the maximum variation obtained by
choosing a different angular window to select the evaporative component,
or by changing the extrapolation to obtain the whole angular distribution,
or by switching on/off the recoil corrections.
More details are given in Appendix B.

Let us first consider the average multiplicities for the 
evaporative emission, $\mathcal{M}_{\mathrm{evap}}$. 
They all display a monotonic increase with increasing TKEL: steeper for 
the most peripheral collisions, flatter for the less peripheral ones. 
However, the magnitude of this increase is different for the different 
particle species. 
In fact, at small TKEL, the average multiplicities range from a few 
tenths for protons down to about 10$^{-3}$ for tritons or even 10$^{-4}$ 
for IMFs, while at the highest TKEL values considered here the 
multiplicities are much more leveled off, with 2--3 protons or $\alpha$ 
particles, as compared to 0.2--0.3 tritons or IMFs.
Thus, the entire evolution with TKEL may be comprised within 
about a decade (protons in the Nb+Nb system), or it may span more than 
three orders of magnitude (IMFs).

Concerning the dependence on the system size, Fig.~\ref{fig:moltepl}
shows that, at a given TKEL, the evaporative multiplicities in the
$^{116}$Sn+$^{116}$Sn collision (open symbols) are systematically s
maller than the corresponding multiplicities in the 
$^{93}$Nb+$^{93}$Nb collision (full symbols). 
This holds true for all particle species and at all TKEL values, 
but it is more clearly visible in the panels for $\alpha$ particles 
and protons and generally in the lowest TKEL bins.
This effect is well reproduced by calculations 
with the statistical code \textsc{Gemini} \cite{Charity88b} and 
may be explained by the fact that 
nuclei produced in the $^{116}$Sn + $^{116}$Sn reaction, because of 
the slightly larger initial value of N/Z (1.32 versus 1.27 of 93Nb),
get rid of their excitation energy 
by emitting more neutrons and less charged particles.

As for the beam-energy dependence, at a given TKEL there is a limited 
sensitivity of $\mathcal{M}_{\mathrm{evap}}$ to this parameter,
but nonetheless one can clearly observe, for all particle species, 
the tendency to rise with decreasing beam energy, a feature that 
becomes more prominent for heavier particles.
For example, in Nb+Nb, the multiplicities are nearly the same for 
protons at all four bombarding energies, while for $\alpha$ particles 
and IMFs at small TKEL they rise by about a factor of 1.5--2 from one 
beam-energy to the next lower one.
A possible explanation is that, as already noted, with increasing beam 
energy the midvelocity emissions draw off an increasing amount of mass 
and energy from the collision, so that a given TKEL corresponds to 
smaller excitations of the primary PLFs and hence to lower values 
of $\mathcal{M}_{\mathrm{evap}}$.
Moreover, at high bombarding energies the contact time is shorter and 
the angular momentum transfer is likely to be smaller so that -- in 
spite of the larger values of the orbital angular momenta in the entrance 
channel -- the primary PLFs may reach lower spin values, an additional 
fact hindering the evaporation of heavier particles.

 \begin{figure}[t]
  \includegraphics[width=80mm,bb=10 10 515 800,clip]{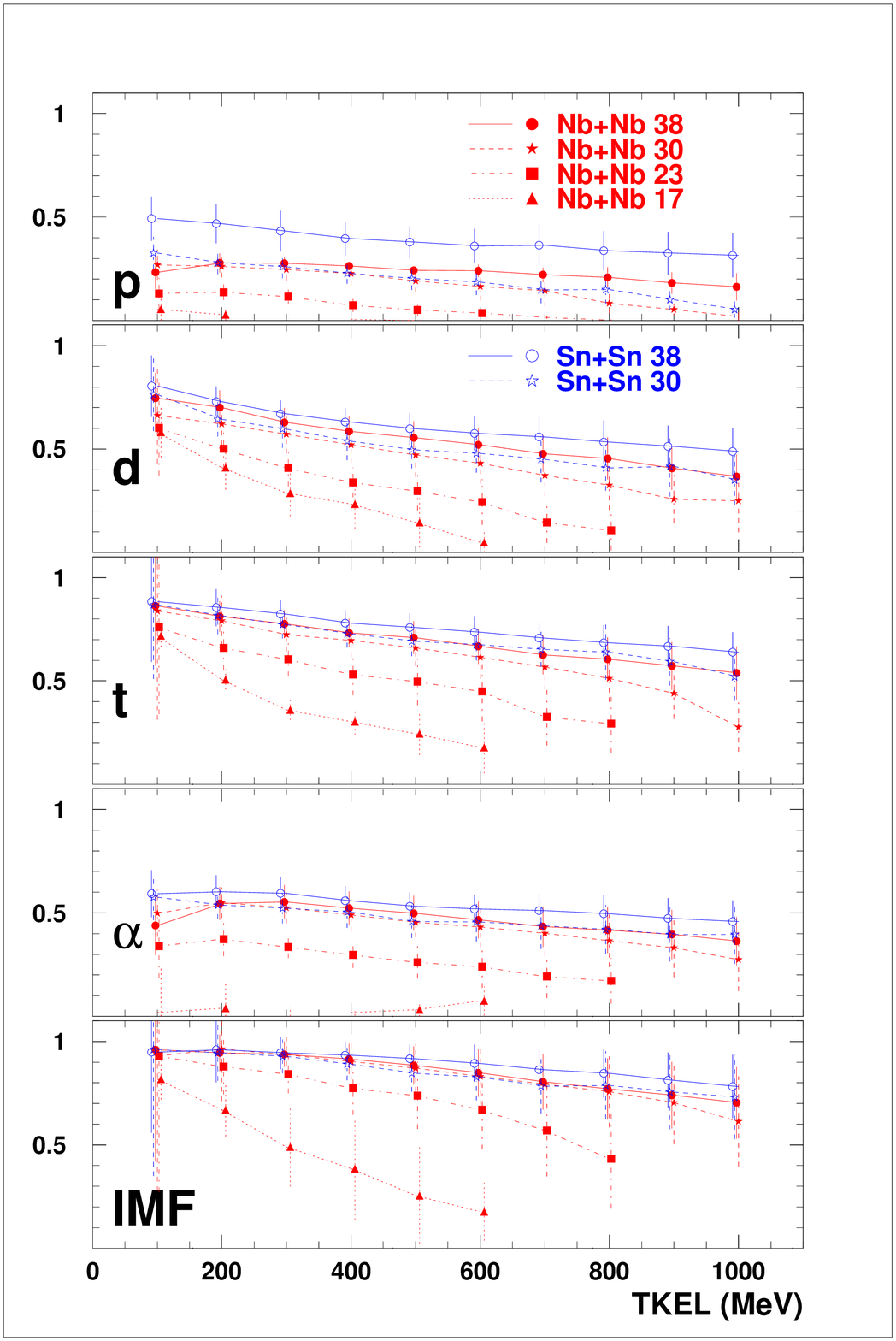}
  \caption{\label{fig:rapp}
  (color online)
   Ratios of midvelocity-to-total multiplicities of forward-emitted 
   p, d, t, $\alpha$ and IMFs, as a function of TKEL.
   Symbols have the same meaning as in Fig. \ref{fig:multot}.
  }
 \end{figure}

In any case, the variations of $\mathcal{M}_{\mathrm{evap}}$ with 
bombarding energy are small when compared with the variations (about two 
orders of magnitude, or more) occurring over the full range of TKEL.  
Thus, at Fermi energies the variable ``TKEL'' is still reasonably well 
correlated with the total amount of particles emitted from the PLF.
This observation supports the conclusion \cite{Mangiarotti03} that,
at these beam energies, equal values of TKEL still indicate comparable 
excitation energies deposited in the PLF (and in the TLF as well).

%
The midvelocity multiplicities $\mathcal{M}_{\mathrm{midv}}$, 
displayed in the right part of Fig. \ref{fig:moltepl},
present an increase with TKEL similar, at first sight, to 
that of the evaporative multiplicities $\mathcal{M}_{\mathrm{evap}}$.
However, a closer inspection shows that their behavior is in many respects 
different and to a certain extent complementary.
First of all there is a less pronounced rise, the entire evolution with TKEL 
spanning --for all particles-- less than two decades;
in some cases the multiplicities seem even to decrease at the highest TKEL.
Then, the dependence on the mass size favors the heavier 
$^{116}$Sn+$^{116}$Sn system, where, at a given TKEL, one generally 
observes larger $\mathcal{M}_{\mathrm{midv}}$ values, 
a feature particularly evident for hydrogen isotopes.
Finally, at variance with what is observed for $\mathcal{M}_{\mathrm{evap}}$, 
at a given TKEL value the midvelocity multiplicities
show an appreciable increase with increasing beam energy,
especially for what concerns the lighter particles.  

In order to better appreciate the different dependence of the evaporative
and midvelocity multiplicities on the beam energy and TKEL, 
Fig. \ref{fig:rapp} presents the ratios of the midvelocity components 
to the total multiplicities measured in the collisions 
$^{93}$Nb + $^{93}$Nb and $^{116}$Sn + $^{116}$Sn.
One observes that generally, for a given beam energy, all ratios 
decrease with increasing TKEL. 
Moreover, with decreasing beam energy, the ratios have smaller values 
and display a faster decrease with increasing TKEL.
With the notable exception of the protons and -- to a lesser extent -- 
of the $\alpha$ particles, at low TKEL
(i.e., in the most peripheral events) the multiplicity ratios
tend to reach values as large as 0.8--0.9.
This fact confirms that peripheral collisions are the best 
environment to investigate the phenomenon of the midvelocity emissions
with the least contamination from sequential evaporation.
This prevalence of the midvelocity particles decreases with increasing
centrality of the collision.
Comparing the different panels, one sees that protons are mainly 
produced in evaporative processes and $\alpha$ particles are almost 
evenly shared between the two mechanisms. 
On the contrary, the most ``exotic'' products, i.e., tritons and 
especially the IMFs, are those which are most specific of the midvelocity 
emissions: at 38\AMeV, even at the highest TKEL, more than 70\% 
of all emitted IMFs are attributed to the midvelocity component 
and this percentage rises to almost 100\% at small TKEL.

 \begin{figure}[t]
  \includegraphics[width=80mm,bb=0 0 567 535]{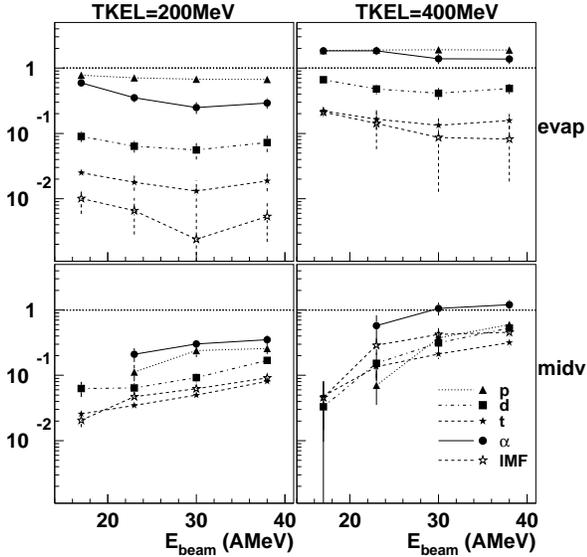}
  \caption{\label{fig:ebeam}
   Dependence on beam energy of evaporative (upper panels) and midvelocity 
   multiplicities (lower panels) for forward-emitted p, d, t, $\alpha$ 
   and IMFs ($Z$=3--7) in the  $^{93}$Nb+$^{93}$Nb collision.
   The data correspond to TKEL values of 200 and 400 MeV
   (left and right panels, respectively). 
  }
 \end{figure}

The different evolution of $\mathcal{M}_{\mathrm{evap}}$ and
$\mathcal{M}_{\mathrm{midv}}$ with bombarding energy is shown in 
Fig. \ref{fig:ebeam} for two TKEL bins in the collision $^{93}$Nb + $^{93}$Nb.
While all evaporative multiplicities stay almost constant, or even 
show a weak decrease with increasing bombarding energy, the midvelocity 
multiplicities display a general increasing behavior, more 
pronounced for the data at higher TKEL. 


\subsection{Nature of the emissions    \label{subs-sqrte}} 

 \begin{figure}[t]
  \includegraphics[width=85mm,bb=0 0 547 520,clip]{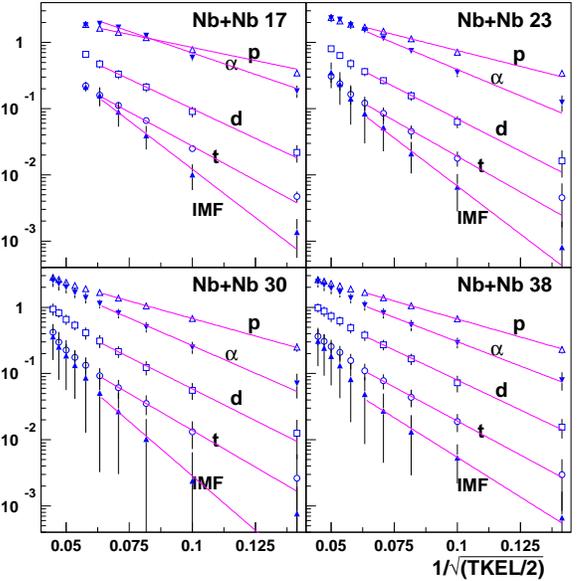}
  \caption{\label{fig:radice}
   (color online)   Multiplicities of PLF-emitted particles
   as a function of $1/\sqrt{\mathrm{TKEL}/2}$.
   for $^{93}$Nb+$^{93}$Nb at 17, 23, 30 and 38\AMeV. 
   Lines are linear fits to the logarithm of the multiplicities in 
   the low-energy part of the excitation functions.
  }
 \end{figure}

In spite of repeated experimental and theoretical efforts over the 
years, the question about the nature of the mechanism(s) responsible 
for the observed LCP and IMF emission is still not completely settled.

For what concerns the PLF emissions, it is natural to expect that 
an evaporation-like de-excitation should be present.
Indeed the usual procedure adopted in the literature 
(see, e.g., \cite{Plagnol99}), and also
in the present paper, for separating 
the two emission components is based on that working hypothesis.
Its validity can be checked by looking at several characteristics of 
this emission, in comparison with the results of calculations with a 
statistical evaporation code like {\sc Gemini} \cite{Charity88b}.
Hereafter we present some pieces of information 
which support the hypothesis.

According to statistical theory, 
the partial decay width associated with a given exit channel should 
present an exponential dependence on the temperature T of the source 
(Boltzmann factor, mainly due to the increase of level density with 
increasing T) of the type: $\Gamma \propto \exp(-{\mathrm{B/T}})$, 
where B is the barrier associated with that channel.
Therefore also the average multiplicities $\mathcal{M}$ of the various 
evaporated particles may be expected to present a similar exponential 
dependence on 1/T, or - in a Fermi gas model - on the inverse square 
root of the excitation energy,
$\mathcal{M} \propto \exp(-{\mathrm{c/\sqrt{E^\star}}})$,
where $c$ is a constant, dependent on the particle species.
In order to compare data with these expectations, the first non-trivial 
task is to estimate the PLF excitation energy.
At low bombarding energies (where midvelocity emissions are negligible 
and TKEL represents the total excitation of the system) and for symmetric 
systems, TKEL/2 is certainly a good average estimate of the excitation 
energy of each of the two primary products of a binary reaction. 
As already noted, with increasing bombarding energy this interpretation 
becomes more and more questionable due to the increasing relevance of 
the midvelocity emissions.
However, the overall energy balance of the reaction $^{93}$Nb+$^{93}$Nb
at 38\AMeV\ performed in Ref. \cite{Mangiarotti04} has demonstrated the 
persistence of an approximately linear correlation between TKEL and the 
excitation energy of the two primary products.
Thus one can suppose that TKEL/2 still represents a measure of the 
excitation energy of the PLF (and of the TLF as well), although it may be 
that the scale is no longer a quantitative one, differing at the various 
bombarding energies.
In Fig. \ref{fig:radice}, the data of the reaction $^{93}$Nb+$^{93}$Nb at
the four bombarding energies of 17, 23, 30 and 38\AMeV\ are 
presented \cite{Mangiarotti03}
in a semilogarithmic plot as a function of $1/\sqrt{TKEL/2}$. 
Indeed, the logarithms of the multiplicities of all particles display 
the same behavior at all bombarding energies, namely a linear one.
This fact is put into evidence by the lines, which are fits to the 
points in the low TKEL region, for each particle species.
 
 \begin{figure}[t]
  \includegraphics[width=85mm,bb=10 10 480 230,clip]{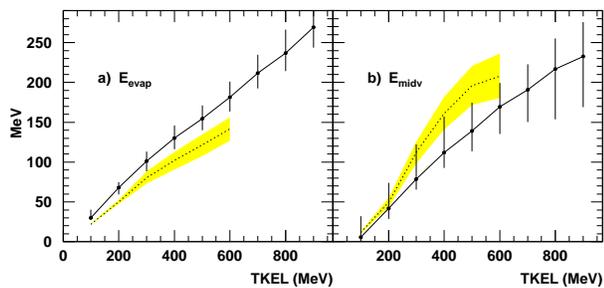}
  \caption{\label{fig:ebil}
   (color online)
   Average excitation energy of PLF (left) and midvelocity ``source'' 
   (right) as a function of TKEL for $^{93}$Nb+$^{93}$Nb at 38\AMeV.
   Uncertainties due to different analyses and assumptions about the 
   n-richness of the midvelocity emissions are indicated by error bars.
   Dashed lines and shaded areas show estimates and uncertainties  
   of Ref. \cite{Mangiarotti04}.
  }
 \end{figure}

A more quantitative scale of excitation energy can be estimated
following the analysis performed in Ref. \cite{Mangiarotti04}
for the reaction $^{93}$Nb+$^{93}$Nb at 38\AMeV.
There the average excitation energy of the primary PLF and of a 
hypothetical midvelocity ``source'' was obtained 
from the total energy balance of the reaction, i.e.,
by summing up the measured kinetic energies of the respective emitted 
particles and taking into account the average Q-value for 
disassembling the system into the final reaction products.
For the semiperipheral events of $^{93}$Nb+$^{93}$Nb at 38\AMeV,
the excitation energies of the two sources have been re-evaluated 
with some improvements in the analysis and the results are shown 
in Fig. \ref{fig:ebil}.
The previous estimates of \cite{Mangiarotti04} (dashed lines) are
indicated too, together with their uncertainties (shaded areas)
resulting from two rather extreme assumptions about the n-richness of 
the midvelocity emissions (from N/Z= 1.1 to 1.44).
For the present evaluation, the uncertainties due to the same hypotheses 
on N/Z and to possible variations in the analysis (see Appendix B) 
are indicated by error bars.
Most of the difference with respect to the previous results 
of \cite{Mangiarotti04} is due to
the use of relativistic kinematics and to the correction for recoil effects.

\begin{figure}[t]
  \includegraphics[width=42mm,bb=30 15 260 260,clip]{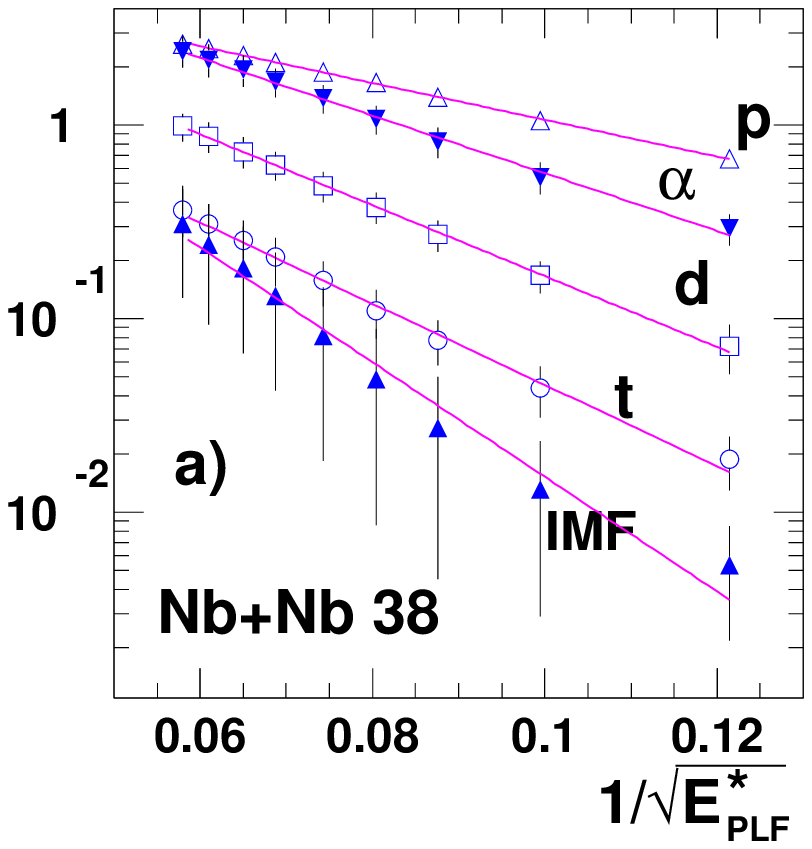}
  \includegraphics[width=42mm,bb=30 15 260 260,clip]{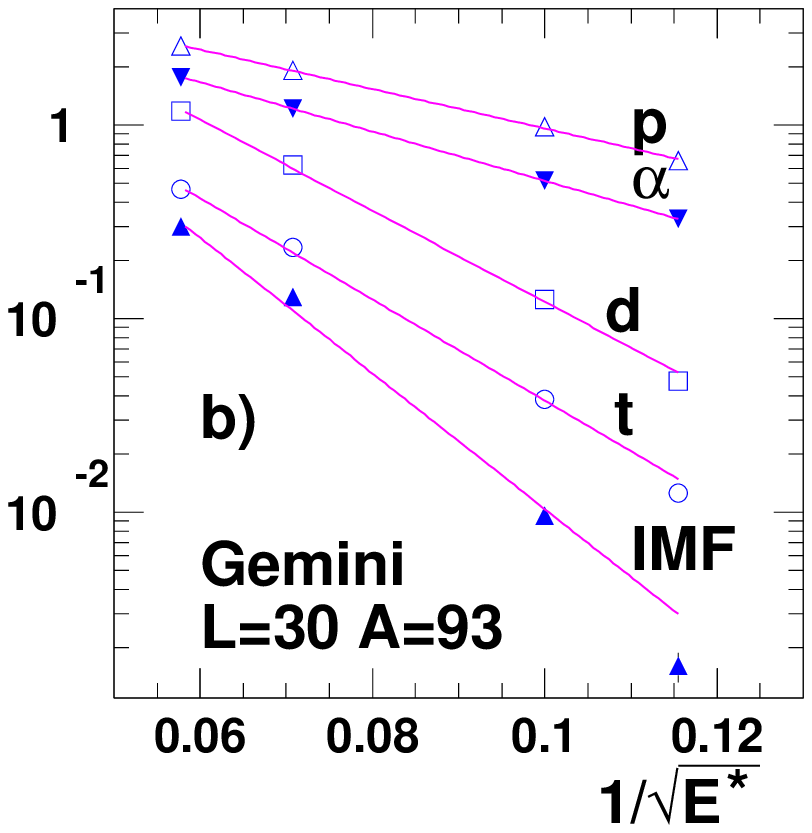}
  \caption{\label{fig:radice2}
   (color online)
   (a) Multiplicities of PLF-emitted particles
   as a function of $1/\sqrt{{\mathrm{E}}^{\star}_{\mathrm{PLF}}}$,
   for $^{93}$Nb+$^{93}$Nb at 38\AMeV,
   with E$^{\star}_{\mathrm{PLF}}$ excitation energy (in MeV) 
   of the primary PLF (symbols as in Fig. \ref{fig:radice}); 
   lines are linear fits to the points.
   (b) Similar plot, for the decay of a $^{93}$Nb nucleus 
   according to \textsc{Gemini} calculations.
  }
 \end{figure}

Figure \ref{fig:radice2}(a) shows 
the PLF particle multiplicities as a
function of the inverse square root of the excitation energy
of the primary PLF, $1/\sqrt{E^\star_{\mathrm{PLF}}}$.
[The first point (TKEL = 100 MeV) is not used here
 because of the large relative uncertainty of its position 
 on the x-axis, depending on the analysis method (see Appendix B).]
When this more appropriate estimate of the excitation energy 
of the ``source'' is used, an improved linear behavior is apparent in 
the logarithmic presentation, as shown by the displayed linear fits to 
the points.
Thus the observed shape of the excitation functions is compatible with a 
process ruled by statistical laws.

\begin{figure}[t]
  \includegraphics[width=90mm]{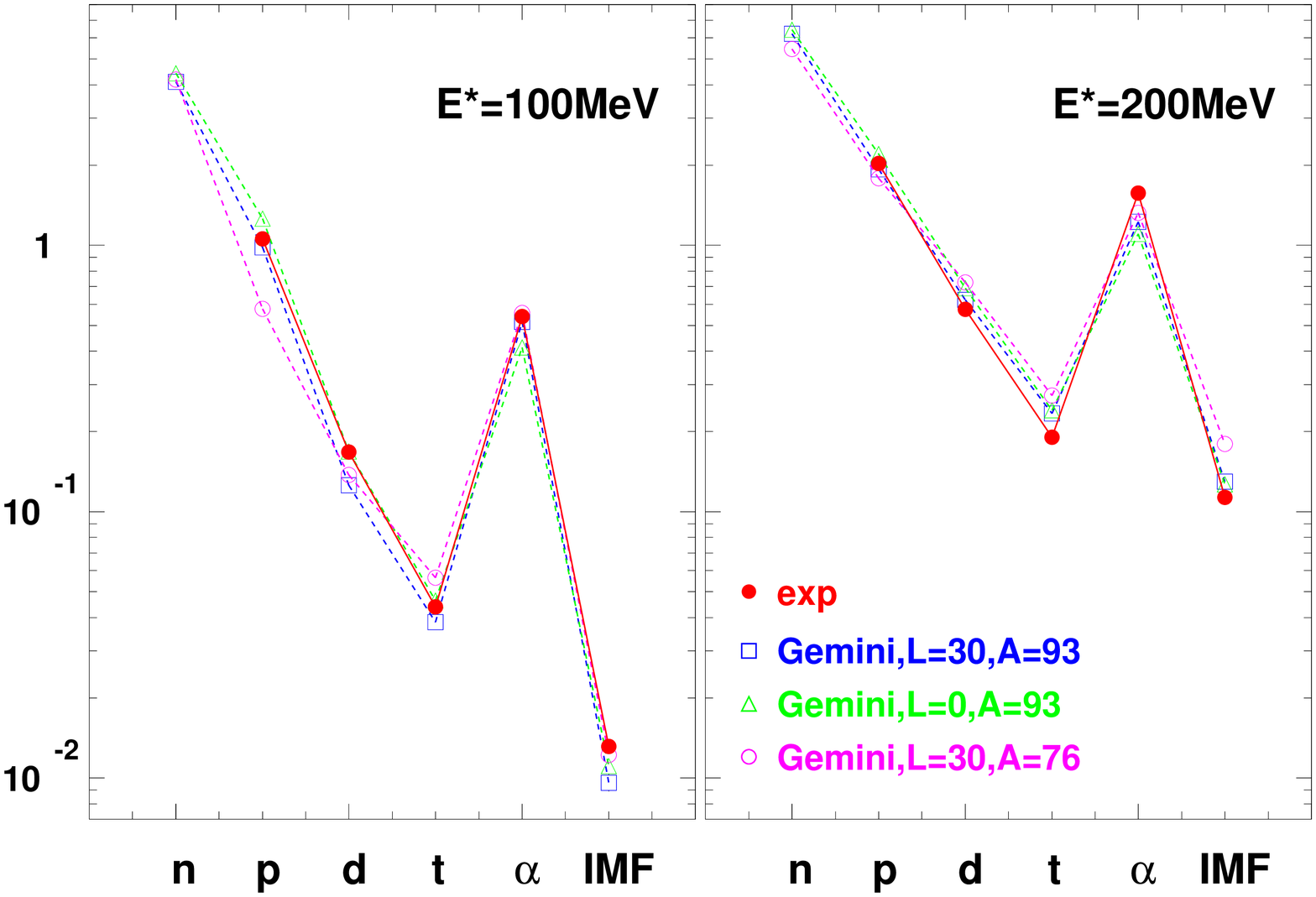}
  \caption{\label{fig:gemLCP}
   (color online)
   Multiplicities of light particles and light IMFs (Z=3--7)
   evaporated from the PLF in
   $^{93}$Nb+$^{93}$Nb at 38\AMeV, for an estimated PLF excitation energy 
   of 100 and 200 MeV (left and right panel, respectively).
   Experimental data (full dots) are compared with \textsc{Gemini} 
   calculations (open symbols) with different mass and 
   spin of the source; for the calculations also the 
   neutron multiplicities are shown.
  }
 \end{figure}

A similar plot, based on the multiplicities calculated with the 
code \textsc{Gemini} for the decay of an excited $^{93}$Nb 
of spin 30~$\hbar$, is shown in Fig. \ref{fig:radice2}(b).
The agreement of the calculations with the experimental data for the 
PLF-emissions is rather good, thus confirming the evaporative origin 
of this component.
No attempt has been made to take into account, in a detailed way, 
possible changes of mass and spin of the primary PLF with increasing 
E$^\star$, as this was beyond the scope of the comparison.
However the relatively weak sensitivity of the calculations to these 
parameters and the good agreement with the experimental data 
for the PLF emissions are demonstrated in Fig. \ref{fig:gemLCP}. 
In this picture the open symbols show the multiplicities 
of n, p, d, t, $\alpha$ particles and light IMFs ($Z$=3--7) 
calculated with the {\sc Gemini} code
for two values (E$^\star$= 100 and 200 MeV) of the excitation energy of 
the evaporating source and for different assumptions on its mass and spin. 
The full dots show the corresponding experimental 
multiplicities of PLF-emitted charged particles (neutrons are not 
measured) for two TKEL bins which, according to the energy balance 
of Fig. \ref{fig:ebil}(a), correspond to the selected values of E$^\star$.
Again, the good agreement between experimental data and calculations 
indicates that the decay of PLFs produced in peripheral and 
semiperipheral collisions is compatible with the usual evaporation of 
an excited nucleus at normal density, at least for what concerns the 
first 3--4 fm of overlap (see Appendix A).
So, at variance with the claims of other authors \cite{Milazzo02},
in order to reproduce our data there is no need to resort
to models (like SMM \cite{Bondorf95}) which describe the fragmentation 
of an expanding diluted nucleus (typically at one third of normal density).
Possibly, the reason is that the events analyzed by the cited authors 
are less peripheral (indeed their estimated PLF excitation energies 
are around 4--5\AMeV, i.e., higher than ours) and 
correspond to a rather particular event selection 
(at least 3 IMFs) and not to the bulk of the collisions as in our case.
It has also been verified that, as expected for an 
equilibrated emission, the velocity spectra of the emitted LCPs have 
approximately Maxwellian shapes with slope parameters of the order of 
2--5 MeV, depending on the selected TKEL value.

\begin{figure}[t]
  \includegraphics[width=70mm]{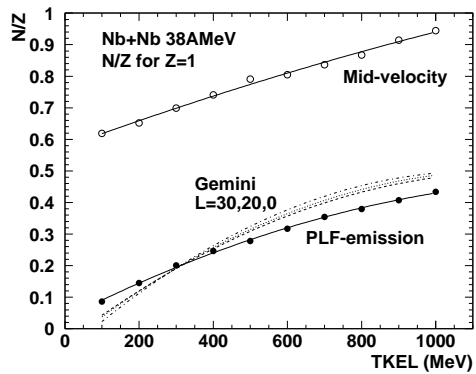}
  \caption{\label{fig:gem_nsuz}
   Average value of N/Z for evaporative (full dots) and midvelocity 
   (open dots) emission of Hydrogen as a function of TKEL in 
   $^{93}$Nb+$^{93}$Nb at 38\AMeV.
   Dashed lines are the results of calculations with the 
   code \textsc{Gemini}.
  }
 \end{figure}

Finally, the lower part of Fig. \ref{fig:gem_nsuz} shows 
the average N/Z ratio of hydrogen isotopes emitted by the PLF in the 
reaction $^{93}$Nb+$^{93}$Nb at 38\AMeV\ (full dots).
This ratio increases with TKEL from 0.1 to about 0.4 and is compatible with 
the results of \textsc{Gemini} calculations for the decay of a $^{93}$Nb 
nucleus at normal density with the appropriate excitation energy
obtained from Fig. \ref{fig:ebil}(a).


For what concerns the midvelocity component, some characteristic aspects 
(like, e.g., the emission pattern of LCPs and IMFs, the space-time 
extension of the ``source'' and its associated energy
\cite{Montoya94,Plagnol99,Piantelli02,Defilippo05neck,Hudan05})
have been already investigated in the past.
However, very different hypotheses have been proposed for the 
underlying mechanism, ranging from fully dynamical processes 
(like, e.g., surface instabilities in the non-spherical, transient shapes 
of the interacting system \cite{Baran04}) to purely statistical ones (like, 
e.g., a proximity-enhanced statistical decay of PLF and TLF in the external 
inhomogeneous Coulomb field of the other reaction 
partner \cite{Botvina99,Botvina01}),
although it cannot be excluded that more than one mechanism contributes to 
the observed phenomena \cite{Piantelli02}.

The average N/Z ratio of the hydrogen isotopes
 in the midvelocity component is shown by the open dots in the upper part 
 of Fig. \ref{fig:gem_nsuz}, again for the reaction $^{93}$Nb+$^{93}$Nb 
 at 38\AMeV. 
 This ratio has now much larger values (from 0.6 to about 1.0),
 thus indicating a substantial difference between the emissions from the 
 PLF and those at midvelocity.
 This may be an indication of neutron enrichment at midvelocity, 
 as proposed on the basis of the emission of complex particles
 \cite{Dempsey96,Lukasik97,Plagnol99,Larochelle00,Milazzo02,Piantelli02},
 or it may be somehow related to the reduced size of 
 the ``source'' and hence to its higher energy concentration, 
 as indicated in \cite{Mangiarotti04}.

\begin{figure}[t]
  \includegraphics[width=42mm,bb=30 15 260 260,clip]{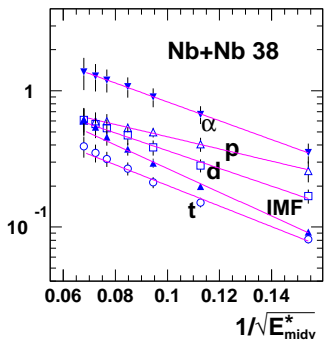}  
  \caption{\label{fig:radice2midv}
   (color online)
   Semilogarithmic plot of midvelocity particle multiplicities
   as a function of $1/\sqrt{E^{\star}_\mathrm{midv}}$
   for Nb+Nb at 38\AMeV\ 
   (symbols as in Fig. \ref{fig:radice}).
   $E^{\star}_\mathrm{midv}$ is the excitation energy of the emitting 
   ``source'', lines are fits to the points.
  }
 \end{figure}

The energy deposited in the matter at midvelocity, $E^\star_\mathrm{midv}$, 
estimated from the energy balance of Fig. \ref{fig:ebil}(b),
is used in Fig. \ref{fig:radice2midv}
to present also the midvelocity multiplicities\footnote{Here, 
          as in Ref. \cite{Mangiarotti04}, the particle 
          multiplicities and $E^\star_\mathrm{midv}$ of the 
          midvelocity ``source'' refer to the forward-going 
          particles only; for the total midvelocity processes 
          one has to double both the multiplicities 
          and the estimated $E^\star_\mathrm{midv}$.}
in a logarithmic plot as a function 
of $1/\sqrt{E^\star_\mathrm{midv}}$:
to our knowledge, this is the first time that 
midvelocity multiplicities are presented in this way.
(Here too, the first point at TKEL = 100 MeV has not been used.)
First of all, we want 
to point out some remarkable differences between the midvelocity 
emissions of Fig. \ref{fig:radice2midv} and the PLF emissions of 
Fig. \ref{fig:radice2}(a).
For example, 
all midvelocity multiplicities are compressed within about one decade 
(note the different horizontal and especially vertical scales of the 
two figures) and consequently their slopes are all sizably flatter.
Even more striking is the fact that the relative abundances of the 
emitted particles are different, with remarkable inversions between 
protons and $\alpha$ particles and between tritons and IMFs.

In this presentation as a function of $1/\sqrt{E^{\star}_\mathrm{midv}}$, 
 one observes a linear correlation
 also for the midvelocity emissions, as demonstrated by the linear fits 
 to the points, but no easy interpretation is at hand.
For a different process, namely multifragmentation in central collisions, 
it was argued \cite{Moretto93} that such a linear behavior (called 
thermal scaling) would be the proof of the statistical, sequential 
decay of the ``source'' (the role of dynamics being relegated to its 
formation), but this conclusion is amply disputed. 
In fact, very different models proved able to reproduce a similar 
behavior, and no general consensus is
reached on aspects like the appropriate treatment of 
the data and estimation of the source temperature, the sequentiality or 
simultaneity of the process, the role of fluctuations and correlations 
(see, e.g., \cite{Botvina95,Elliott00,Beaulieu01} and references therein).
Although the linear behavior in itself may be suggestive of
a thermal process, its mere observation is not a proof and one actually
expects that in peripheral collisions the dynamics should play a much 
more important role than in central collisions.

As a last point,
it is worth noting that these striking differences 
between evaporative and midvelocity multiplicities
give strong support to and demonstrate the effectiveness of the 
separation procedure outlined in Sec. \ref{subs-statmidv}, which has 
been used to disentangle the two components.

\section{Summary and Conclusions \label{sec:concl}}

The symmetric collisions $^{93}$Nb + $^{93}$Nb at 17, 23, 30 and 
38\AMeV\ and $^{116}$Sn+$^{116}$Sn at 30 and 38\AMeV\ have been 
investigated using the \textsc{Fiasco} setup, a low-threshold 
multidetector covering a large fraction of the forward solid angle.

The present analysis is focused on peripheral and semiperipheral 
events resulting from binary or quasi-binary reaction processes. 
They are characterized by the presence in the exit channel of two heavy 
remnants, which can be identified as the projectile- and target-like 
fragments.
These binary or quasi-binary processes are found to exhaust more than 
half of the expected total reaction cross section at all the investigated
beam energies.

The kinematic variable TKEL has been used for selecting event samples 
corresponding, on average, to decreasing values of the impact parameter.
It is worth stressing that TKEL, which at low bombarding energies
represents a good estimate of the total excitation of the colliding nuclei,
looses such a physical meaning at Fermi energies and therefore it is 
used here only as an ordering parameter.

The average multiplicities $\mathcal{M}$ of light charged particles 
and intermediate mass fragments with $3 \leq Z \leq 7$ have been 
obtained as a function of TKEL for all the investigated systems.
The data have been evaluated in a range of TKEL values which is 
estimated to roughly correspond to the outermost 30\% of the 
impact parameter range leading to nuclear interaction.
The emission pattern of these particles in the  
($v_{\,\parallel}$,$v_{\,\perp}$) plane -- the velocity components being 
referred to the PLF - TLF separation axis in the c.m. reference system --
clearly shows the presence of two components, one representing
an emission from the excited PLF (and TLF) and a second one at midvelocity, 
i.e., at velocities intermediate between that of PLF and TLF.
A careful analysis of the shape of the angular distribution in the 
reference frame of the PLF has been applied in order to distinguish and 
estimate the multiplicities of particles emitted by the PLF and, by a 
subtraction procedure, those of the particles emitted in the 
midvelocity region.
Correction have been applied to take into account the efficiency of the 
setup, as well as physical effects due to spin and recoil of the
emitting nuclei.

Both emission components increase with increasing TKEL (i.e., 
decreasing impact parameter), although their behavior is different.
More ``exotic'' particles (like tritons, IMFs and to a lesser extent 
deuterons) are characteristic of midvelocity processes in the most
peripheral events, where they outnumber the emissions from PLF.
Moreover, for a fixed TKEL, the midvelocity emissions present a 
clear increase with increasing beam energy, while the emissions from 
PLF show little dependence on the beam energy.

The dependence of PLF emissions on the estimated excitation energy 
of the PLF follows that expected for a decay mechanism governed by a 
barrier and is well described within statistical models.
In fact, calculations with the statistical code \textsc{Gemini} well 
reproduce all experimental features, including the slopes of the 
excitation function, the relative and absolute abundances of the 
various particle species and also the isotopic composition of 
Hydrogen particles.
Thus, at least for the peripheral and semiperipheral collisions, 
the decay of the PLF (and TLF) is in many aspects compatible with the 
usual evaporation from an excited nucleus at normal density.

Surprisingly, also the midvelocity emissions display a 
similar type of dependence on the amount of energy which 
is localized in the hypothetical midvelocity ``source''.
However, this mechanism is quite different from the usual evaporation, 
as it is shown by the inversion in the relative abundance of the emitted 
particles and by the tendency to emit more n-rich light particles.
Indeed, the emission pattern observed in this work and in previous 
ones \cite{Piantelli02}, clearly demonstrates that this hypothetical  
``source'' is not a simple spherical one, sitting at midvelocity and
isotropically emitting particles; 
on the contrary, in order to reproduce the observed emission patterns, 
one needs a more complex ``source'', extended both in space and time.
For a satisfactory explanation of its characteristic features,
one probably needs a dynamic description of the collision 
\cite{Baran04}, but its successive decay seems to possess some features 
reminiscent of a statistical process.

\begin{acknowledgments}

We are very grateful to A. Gobbi for many stimulating discussions.
Many thanks are due to L. Calabretta, D. Rifuggiato and coworkers for 
delivering pulsed beams with excellent timing, to the staff of LNS for 
continuous support, to C. Marchetta for providing targets of very 
good quality, 
     and to the {\sc Hodo-Ct} group for the
     collaboration during the data taking.
Thanks are also due to P. Del Carmine, M. Ciaranfi, G. Tobia and to 
the personnel of the mechanical workshop in Florence for invaluable 
help in the preparation of the experiment.

\end{acknowledgments}

\appendix

\section{Impact parameter estimation
                                                      \label{sec:appA}}

Various experimental variables are
used in literature for estimating
how central or peripheral a collision is and for sorting the measured 
events in (nearly) homogeneous samples.
Among the variables related only to the main
reaction partners, an often used one is the secondary charge 
$Z_{\mathrm{sec}}$ of the PLF residue (at low bombarding energies or for 
quasi-peripheral events) or the charge of the heaviest fragment 
$Z_{\mathrm{max}}$ (at higher energies or for more central collisions): 
in fact, on average, the lighter this charge, the more
violent and central the collision is likely to be \cite{Djerroud01}.  
Such non-kinematic variables may be used for studying kinematic aspects
of particles (like, e.g., their angular distributions), but they are not
equally well suited for studying multiplicities. 
Because of the finite (and not too large) total charge of the system, 
spurious correlations may appear.

In order to avoid this problem, it is preferable to sort the data according
to a kinematical variable, such as the secondary velocity 
$v^{\;\mathrm{Lab}}_{\;\mathrm{PLF}}$ of the PLF in 
the laboratory reference system  \cite{Yanez03}, 
or the relative velocity $v_{\;\mathrm{rel}}$ of the two main reaction 
products,
or the energy loss per nucleon $\epsilon^\star$ of a binary 
collision \cite{Metivier00}.
In this paper we have adopted TKEL, as defined in Eq. (\ref{eq:TKEL}).
In case of a binary process with frozen initial mass asymmetry one expects
to find a good correspondence between all these kinematic variables,
but if the mass asymmetry in the exit channel is allowed to change, then
TKEL has the advantage of explicitly taking into account these variations.

\begin{figure}[t]
   \includegraphics[width=70mm]{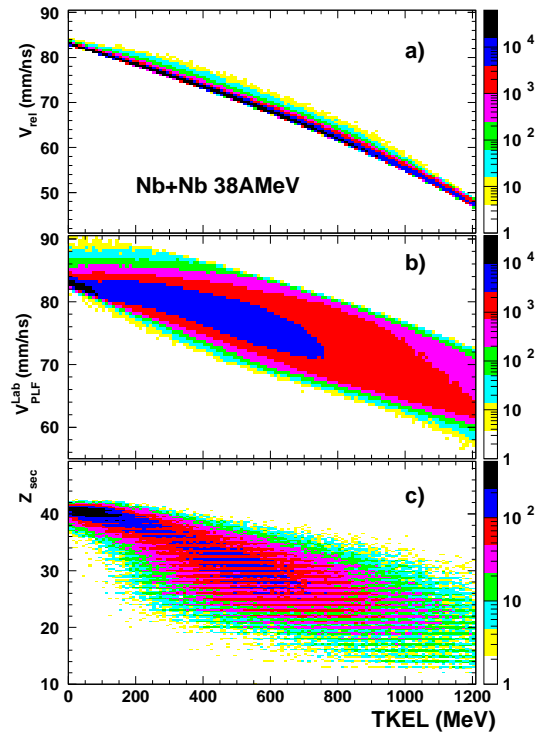}
   \caption
   {\label{fig:tkelcor}
   (color)
   Experimental
   correlations between the kinematic variable TKEL 
   [see Eq. (\ref{eq:TKEL})] and 
   (a) relative PLF-TLF velocity; 
   (b) lab-velocity of the PLF; 
   (c) secondary charge of the PLF. 
   Data refer to $^{93}$Nb + $^{93}$Nb at 38\AMeV.
   }
 \end{figure}

The experimental correlation of TKEL with $v_{\;\mathrm{rel}}$ and 
$v^{\;\mathrm{Lab}}_{\;\mathrm{PLF}}$ for the data of $^{93}$Nb + $^{93}$Nb 
at 38\AMeV\ is shown in Fig. \ref{fig:tkelcor}(a) and (b), respectively. 
The first correlation, TKEL--$v_{\;\mathrm{rel}}$, is extremely narrow 
in the whole considered range:
this is little surprise, since in symmetric collisions most binary exit 
channels have nearly the same reduced mass, so that TKEL is
approximately proportional to $v_{\;\mathrm{rel}}^2$ 
[see Eq. (\ref{eq:TKEL})].
The next correlation, TKEL vs $v^{\;\mathrm{Lab}}_{\;\mathrm{PLF}}$,
is narrow at low TKEL, but tends to become wider with increasing 
inelasticity of the collision.
Finally, Fig. \ref{fig:tkelcor}(c) also shows the correlation of TKEL 
with $Z_{\mathrm{sec}}$: the two variables are well correlated but with 
rather large fluctuations.

\begin{figure}
   \includegraphics[width=70mm,bb=0 0 510 720,clip]{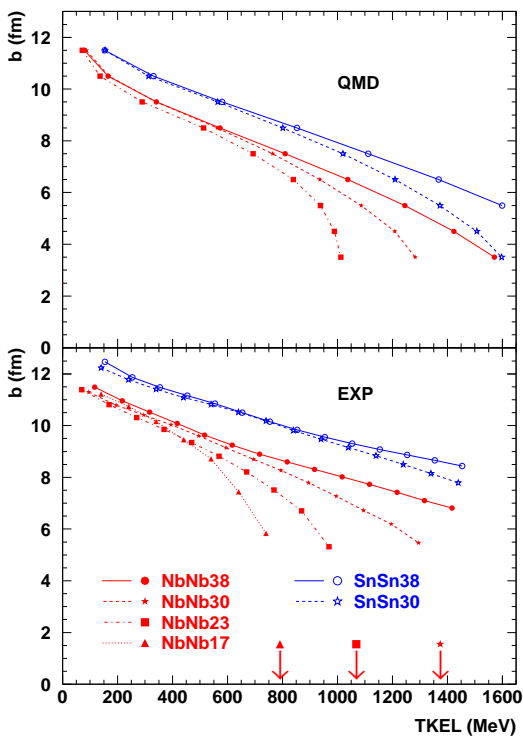}
   \caption
   {\label{fig:parurto}
   (color online)
   Upper panel: average correlation between impact parameter and the 
    kinematical variable TKEL for events generated by the QMD code 
    \textsc{Chimera} \protect{\cite{Lukasik93}}.
   Lower panel: average correlation between impact parameter and TKEL
    obtained from the integration of the experimental reaction cross 
    section for $^{93}$Nb+$^{93}$Nb at 17, 23, 30, 38\AMeV\ and
    $^{116}$Sn+$^{116}$Sn at 30, 38\AMeV;
    arrows indicate the c.m. available energies which are within the 
    displayed energy range. Lines are just an aid to guide the eye.
   }
 \end{figure}

The usage of TKEL as an impact parameter estimator is supported by
the kinematic analysis of events generated with the Quantum Molecular
Dynamics (QMD) code \textsc{Chimera} \cite{Lukasik93}.
The velocity vectors of the two main fragments produced by such QMD 
calculations were used to deduce the variable TKEL, just with the same 
kinematic procedure used for the experimental data: 
the result of this analysis shows a nice correlation
between impact parameter and TKEL \cite{Piantelli01}.   
The average value of the obtained TKEL, plotted as a function of the impact 
parameter used as input to the calculations, is shown in the upper part of 
Fig. \ref{fig:parurto} for the investigated reactions
(except for the 17\AMeV\ case, because at such a low bombarding energy 
the QMD approach seems hardly applicable).
As expected, there is a monotonic increase of TKEL when the impact 
parameter is decreased starting from the
grazing value $b_{\mathrm{graz}}$.
Less obvious is the finding that, for a given system, at large impact 
parameters the correlations are almost independent of the beam energy.
The model suggests that, at different bombarding energies, peripheral 
events sampled in the same TKEL bins may correspond to collisions with 
roughly the same impact parameter.
As the collisions become less and less peripheral, the curves 
progressively separate and bend down, as they approach 
-- at different TKEL values -- the limit of their 
respective available energy in the c.m. system.
Concerning the comparison of the two systems, for $^{116}$Sn + $^{116}$Sn 
the correlations begin at a larger value of b, as might have been expected, 
and proceed almost parallel to those of the 
lighter system $^{93}$Nb + $^{93}$Nb.

As an alternative, one can perform an association of TKEL 
with impact parameter using a method already applied in 
Ref. \cite{Stefanini95} and based on a refinement of 
an even
older recipe \cite{Schroeder78}.
The basic hypothesis is that there is a good average correlation between 
decreasing impact parameter and macroscopic behavior of the reaction 
products in terms of scattering angle and kinetic energy dissipation. 
So one can try to follow the evolution of the reaction from the 
two-dimensional plot $d^{\,2} \sigma/ d(\theta_{c.m.}) d({\mathrm{TKEL}})$, 
the so-called Wilczynski-plot \cite{Wilczynski73}, 
and empirically determine -- on average -- an impact parameter scale by 
means of an integration of the reaction cross section, 
starting from the elastic region, going over to the quasi-elastic region 
and then down into the deeply inelastic region; 
events leading to fusion or to several heavy fragments are 
assumed to be located at the lower end of the impact parameter scale.

In practice, during our experiment, events from a minimum bias trigger 
(``singles'', requiring a number M$\geq$1 of hits in the gas detectors)
were acquired at a reduced rate, together with rarer events 
from more selective triggers (requiring higher-fold hits).
For the four most forward gas detectors, ``clean'' angular distributions 
can be obtained for these minimum-bias events by selecting appropriate 
equal windows in the azimuthal angle and further requiring that the 
time-of-flight be compatible with that of elastically scattered projectiles.
The so obtained angular distributions nicely reproduce the
$1 / \sin^4(\theta_{c.m.}/2)$ shape expected for the Rutherford scattering 
of pointlike charges, until the region near the grazing angle is reached.
The method is very sensitive; small differences in the rates of the 
four most forward gas detectors, have been attributed to a misalignment 
(generally less than one tenth of a degree) of the beam with respect to the 
optical axis and were used to correct the polar angles.
For each of the investigated reactions, a simultaneous fit of the
Rutherford formula to the angular distributions of the four gas detectors 
has been used to estimate the conversion factor from counts to millibarns,
with an uncertainty below a few percent.

\begin{figure}[t]
  \includegraphics[width=75mm]{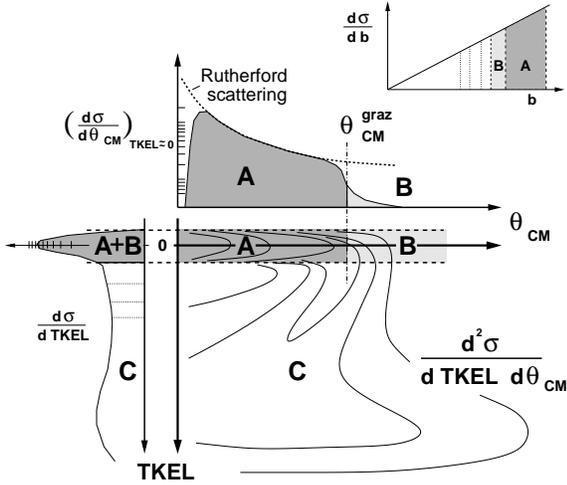}
  \caption
  {\label{fig:wilcz}
   Sketch of the cross section integration
  }
 \end{figure}

The procedure is then illustrated with the help of Fig. \ref{fig:wilcz},
where the correlation 
$d^2 \sigma / d \theta_{\mathrm{c.m.}} d \mathrm{TKEL}$ 
for binary events is sketched together with its total projection 
$d \sigma/ d \mathrm{TKEL}$ (on the left) and a projection 
$d \sigma / d \theta_{\mathrm{\,c.m.}}$ of the elastic and 
quasi-elastic region (on the top).
The region $A$ is dominated by elastic scattering (which is responsible 
for the large peak around TKEL$\approx$0 in the projection on the left),
while regions $B$ (still at TKEL$\approx$0, but beyond the grazing angle 
$\theta^{\mathrm{graz}}_{\mathrm{\,c.m.}}$) and $C$ are populated 
by reactions.
The two-body cross sections of Table \ref{tab2} have been obtained by
integrating the efficiency corrected experimental yields in regions $B$ 
and $C$ (with a small correction to take into account the quasi-elastic
cross section lying below the elastic ridge in region $A$), while
the three-body cross sections come from the lower part of region $C$ only.

The correlation between impact parameter $b$ and TKEL is obtained by 
filling the triangle-shaped distribution $d \sigma / d\, b$ 
(upper part of Fig.~\ref{fig:wilcz})
in successive steps.
Starting from a value of $\theta_{\mathrm{\,c.m.}}$ still in the 
elastic region 
(which is uniquely related to impact parameter by Rutherford scattering),
one integrates the measured cross section at larger angles 
(remaining part of the region $A$ and region $B$), thus determining a new
impact parameter value, which is the first point in the experimental 
correlations [see Fig. \ref{fig:parurto}(b)].
Now one proceeds towards smaller and smaller impact parameters by 
integrating the measured cross section for more and more inelastic events
(in region $C$).
In a ``sharp cut-off'' picture, each successive integrated piece of cross 
section is used to completely fill the next (to the left) slice of the 
triangle $d\sigma/ d\,b$, thus determining the next point of correspondence 
between impact parameter $b$ and TKEL.
Of course, for this procedure to work reasonably, what must be integrated 
is the total reaction cross section.
For the present work, focused on peripheral collisions, 
this means that it is necessary to obtain:
\begin{itemize}
\item[i)]
   a good quantitative measurement of the most 
   representative exit channels of the reaction 
   (i.e., two-body and, to a lesser extent, three-body events);
\item[ii)]
   a good correction for the inefficiencies of the setup
   (``experimental filter'');
\item[iii)] 
   a smooth transition, without detection gaps, from the elastic 
   to the inelastic events
   (i.e., not too complicated triggers and good efficiency for
    quasi-elastic events).
\end{itemize}

The results for the investigated reactions are presented in the lower 
panel of Fig. \ref{fig:parurto}.
Qualitatively there is a good agreement with the results of the QMD 
estimates, as the same main features are well reproduced:
(1) There is the same monotonic decrease of $b$ with increasing TKEL. 
(2) For very peripheral collisions the correlations are almost independent 
of the beam energy, while they tend to separate at smaller impact 
parameters. (3) For $^{116}$Sn + $^{116}$Sn the curves are displaced by 
an almost constant value of about 1 fm to larger impact parameters.

For each curve, the first point -- that at the highest $b$ -- is the most 
critical one and it is estimated to be affected by an overall uncertainty 
of the order of 1 fm (due to corrections of misalignment of the beam axis, 
fitting with Rutherford scattering, integration of elastic cross section, 
matching of elastic and inelastic events).
Being the curves obtained by successive integrations of the cross section,
the errors are strongly correlated and the uncertainty on the first point 
propagates to all other points; in other words, if the first point needed 
to be moved up or down, all the others would move almost rigidly in the 
same direction.
Considering that the \textsc{Fiasco} setup has been developed and 
optimized for peripheral reactions, this impact parameter determination 
becomes less reliable when the collisions become more central
In fact, one may start missing some relevant exit channel and, 
in turn, this progressive underestimation of the total reaction cross 
section may cause the impact parameter to decrease too slowly with TKEL.
Indeed, this might explain why, at large TKEL and especially for larger 
beam energies, the experimental curves shown in the lower panel of 
Fig. \ref{fig:parurto} do not bend downwards as much as the QMD calculations.

In the TKEL ranges used for the evaluation of the multiplicities
presented in this paper (TKEL $\leq$ 600 and 800 MeV for Nb+Nb at 
17 and 23\AMeV, respectively; $\leq$1000 MeV in all remaining cases)
the binary channel accounts for about 2.0--2.2 barn in the $^{93}$Nb + 
$^{93}$Nb collisions and about 1.9 barn in the $^{116}$Sn + $^{116}$Sn 
collisions (corresponding to $\sim$ 40--50\% and $\sim$ 35\% 
of $\sigma_{\mathrm{reac}}^{\mathrm{\,calc}}$, respectively),
with an uncertainty of about 200--300 mb.
The applied restriction on the ratio of the two c.m. velocities 
(0.4 $\leq v^{\mathrm{c.m.}}_{\mathrm{PLF}}/
(v^{\mathrm{c.m.}}_{\mathrm{PLF}}+v^{\mathrm{c.m.}}_{\mathrm{TLF}}) \leq$ 0.6)
cuts less than 5\% of the considered cross section and is 
appreciable only in the last considered TKEL bin.
From the integration of the experimental cross section it is estimated
that the presented multiplicities refer approximately to the outermost 
third of the impact parameter range 
($b^{\mathrm{exp}}$ / $b_{\mathrm{graz}}^{\mathrm{calc}} \geq$ 60--70\%).
In this region, choosing the same TKEL bin allows to compare results 
at roughly the same impact parameter if the system is fixed and only 
the beam energy is varied, while it is confirmed that there is a shift 
of about 1 fm between the estimated impact parameters for the two systems 
$^{93}$Nb + $^{93}$Nb and $^{116}$Sn + $^{116}$Sn.


\section{Dependence on the analysis method
          \label{sec:appB}}

In order to estimate the uncertainties which may affect the 
determination of the average evaporative and midvelocity multiplicities 
($\mathcal{M}_{\mathrm{evap}}$ and $\mathcal{M}_{\mathrm{midv}}$, 
respectively), the data have been analyzed by using slightly different 
procedures for separating the total multiplicities into these two components. 
More specifically, somewhat different choices have been taken at three 
important points of the analysis and the induced variations have been 
considered as representative of the sensitivity of the results to 
different evaluation procedures.
The three points are
\begin{itemize}
\item[a)]
    the angular range (in the PLF frame) used to estimate the 
    evaporative component,
\item[b)]
    the successive extrapolation of the angular distribution to 
    the whole range 0$^\circ$--180$^\circ$,
\item[c)]
    the corrections for recoil effects.
\end{itemize}

As for point (a), besides the adopted range of 0$^\circ$--45$^\circ$, 
a more conservative one (0$^\circ$--30$^\circ$) has been considered,
as well as the range 0$^\circ$--90$^\circ$, which is often used in 
the literature.
As already discussed with regard to Fig. \ref{fig:occhi-cost}, 
this last choice pays the advantage of estimating the whole angular 
distribution by means of a simple reflection around 90$^\circ$ 
[without the extrapolation of point (b)]
with the disadvantage that the evaporative component may be 
contaminated by the tail of the midvelocity emission.

The extrapolation at point (b) requires some hypothesis on the spin 
value of the evaporating source and its misalignment during the decay,
while for a zero-spin source one expects a roughly 
$\cos{\theta_{\mathrm{\;PLF}}}$-shaped  
angular distribution $d\sigma/d\theta_{\mathrm{\;PLF}}$ 
(although recoil effects may somewhat distort it).
So the considered alternatives are an analytical extrapolation with 
the $\cos{\theta_{\mathrm{\;PLF}}}$ shape and a numerical one,
obtained with a Monte Carlo simulation which assumes a 
non-zero spin value (rising with TKEL from 0 $\hbar$ to 40 $\hbar$) 
and takes into account the progressive spin misalignment along
the evaporation chain.

 \begin{figure}
  \includegraphics[width=80mm,bb=10 10 515 800,clip]{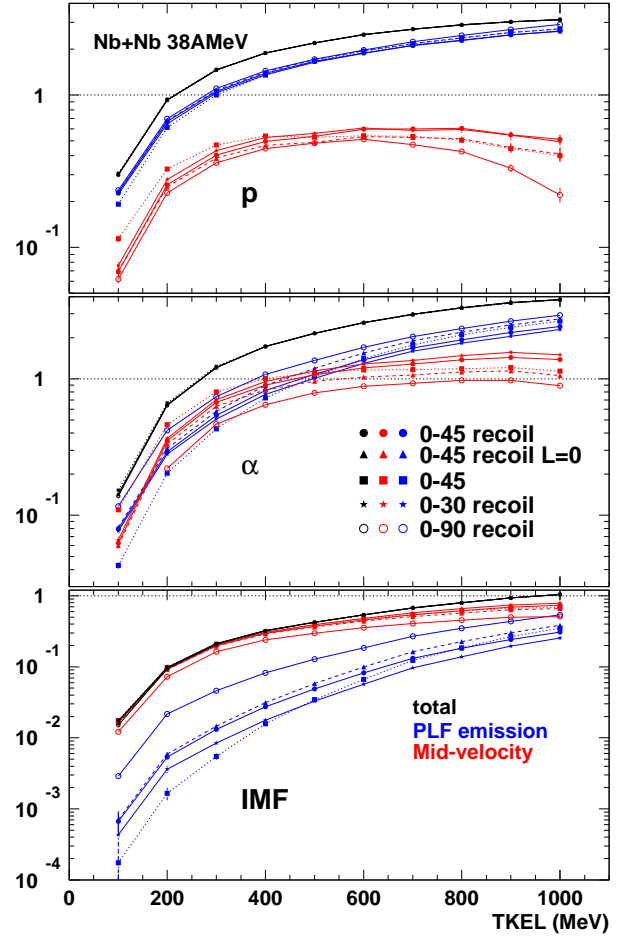}
  \caption{\label{fig:metodi}
   (color) 
   Total (black), evaporative (blue) and midvelocity (red)
   multiplicities for p, $\alpha$ and IMFs 
   (upper, middle and lower panels, respectively) 
   in $^{93}$Nb+$^{93}$Nb at 38\AMeV. The
   results of the analysis adopted in the paper (full dots) 
   are compared with those obtained from slightly different
   analysis procedures (see text).
  }
 \end{figure}

Figure \ref{fig:metodi} shows the differences which are typically 
obtained, by taking as an example the evaporative (blue color) and 
midvelocity (red color) multiplicities of p, $\alpha$ and IMFs 
in the reaction $^{93}$Nb + $^{93}$Nb at 38\AMeV\ 
(the three cases correspond to dominant evaporation, comparable 
components and dominant midvelocity emission, respectively);
the black curves show the corresponding total multiplicities.
The reference (shown by full dots) is always the analysis with the 
choices adopted in this paper, namely evaporation in the angular range 
0$^\circ$--45$^\circ$, extrapolation -- based on Monte Carlo simulations -- 
for non-zero spin source and applied corrections for recoil effects.
Variations are then made one at a time: 
either a zero-spin source is assumed (triangles), 
or the recoil corrections are switched off (squares), 
or the angular range is reduced to 0$^\circ$--30$^\circ$ (stars)
or even enlarged to 0$^\circ$--90$^\circ$ (open dots).

Being the decomposition between $\mathcal{M}_{\mathrm{evap}}$ 
and $\mathcal{M}_{\mathrm{midv}}$ obtained with a subtraction procedure,
these different choices induce larger relative variations in the weaker 
component: thus, in the evaporative component they are largest for the 
IMFs (which are predominantly emitted at midvelocity) and,
in the midvelocity component, they are rather large for protons at 
38\AMeV\ and even larger  -- this time for all particles -- at the 
lowest bombarding energies (where evaporation dominates).

In general one can observe that:
\begin{itemize}
\item 
 The more conservative angular range 0$^\circ$--30$^\circ$ gives values 
 of $\mathcal{M}_{\mathrm{evap}}$ ($\mathcal{M}_{\mathrm{midv}}$) 
 which are just slightly smaller (larger) than those obtained 
 with 0$^\circ$--45$^\circ$.
 Only for IMFs there is a sizable reduction (about -30\%) of the already 
 quite small $\mathcal{M}_{\mathrm{evap}}$ component (while the
 corresponding increase of $\mathcal{M}_{\mathrm{midv}}$ is negligible).
\item
 With the commonly used procedure of estimating the evaporation by taking 
 twice the particles emitted in 0$^\circ$--90$^\circ$, one systematically 
 overestimates $\mathcal{M}_{\mathrm{evap}}$ 
 (and hence underestimates $\mathcal{M}_{\mathrm{midv}}$),
 for all particles and at all bombarding energies. 
 For protons, the variation is negligible on $\mathcal{M}_{\mathrm{evap}}$ 
 and small but sizable on $\mathcal{M}_{\mathrm{midv}}$ 
 (especially at high TKEL); 
 for $\alpha$ the effect is of the order of 30--40\% and comparable on both 
 components;
 for IMFs the effect is huge (about a factor of 2--3) on the weak 
 $\mathcal{M}_{\mathrm{evap}}$ component and much smaller -- but still 
 sizable, about 20--30\% -- on the dominating $\mathcal{M}_{\mathrm{midv}}$.
 Indeed, for IMFs this procedure gives the largest of all considered 
 variations.
\item
 Extrapolating the angular distribution of evaporated particles with a
 $\cos{\theta_{\mathrm{\;PLF}}}$ shape generally overestimates
 $\mathcal{M}_{\mathrm{evap}}$ (and underestimates 
 $\mathcal{M}_{\mathrm{midv}}$).
 However, at small TKEL (where the spin is likely to be small), it produces 
 really negligible variations in all cases.
 At large TKEL, there is a moderate effect (of the order of 10--20\%) 
 only on the weaker component.
\item
 As explained in Sec. \ref{subs-statmidv}, applying no recoil correction
 causes an overestimation of TKEL, namely a given bin erroneously includes 
 some events (which should actually be classified in a bin at lower TKEL) 
 and misses others (which get classified in another bin at larger TKEL).
 The net effect of this gain and loss is determined 
 by the TKEL dependence of the experimental particle yield, i.e.,
 yield of measured events times event multiplicity for that particle). 
 The rise of all multiplicities is steepest at small TKEL and tends 
 to flatten at large TKEL, while the measured event yield doesn't change 
 more than a factor of two over the whole TKEL range.
 Therefore the effect of the recoil correction is largest at small TKEL,
 where it determines an increase of $\mathcal{M}_{\mathrm{evap}}$
 -- of course larger for heavier particles -- from 10--20\% for protons 
 to a factor of about 2 and 3 for $\alpha$ particles and IMFs, respectively
 (the corresponding decrease of $\mathcal{M}_{\mathrm{midv}}$ is about 
 70\% for protons, a factor of 2 for $\alpha$ and negligible for IMFs). 
\end{itemize}
 
 In the Nb+Nb system at 38\AMeV, the effects of the considered analysis 
 variations are most visible in the PLF-multiplicities of IMFs at large 
 beam energies, and in the midvelocity multiplicities of LCPs at low beam 
 energies and large TKEL.
 For all investigated reactions, these effects are schematically 
 represented by the error bars in Figs. \ref{fig:moltepl} and \ref{fig:rapp}
 and they should be kept in mind when comparing the results of different 
 experiments obtained with different analysis procedures.


\bibliography{art_mul}

\begin{thebibliography}{45}
\expandafter\ifx\csname natexlab\endcsname\relax\def\natexlab#1{#1}\fi
\expandafter\ifx\csname bibnamefont\endcsname\relax
  \def\bibnamefont#1{#1}\fi
\expandafter\ifx\csname bibfnamefont\endcsname\relax
  \def\bibfnamefont#1{#1}\fi
\expandafter\ifx\csname citenamefont\endcsname\relax
  \def\citenamefont#1{#1}\fi
\expandafter\ifx\csname url\endcsname\relax
  \def\url#1{\texttt{#1}}\fi
\expandafter\ifx\csname urlprefix\endcsname\relax\def\urlprefix{URL }\fi
\providecommand{\bibinfo}[2]{#2}
\providecommand{\eprint}[2][]{\url{#2}}

\bibitem[{\citenamefont{{{\L}ukasik \textit{et al.}}}(1997)}]{Lukasik97}
\bibinfo{author}{\bibfnamefont{J.}~\bibnamefont{{{\L}ukasik \textit{et al.}}}},
  \bibinfo{journal}{Phys.\ Rev.\ C} \textbf{\bibinfo{volume}{55}},
  \bibinfo{pages}{1906} (\bibinfo{year}{1997}).

\bibitem[{\citenamefont{{Plagnol \textit{et al.}}}(1999)}]{Plagnol99}
\bibinfo{author}{\bibfnamefont{E.}~\bibnamefont{{Plagnol \textit{et al.}}}}
  (\bibinfo{collaboration}{INDRA collaboration}), \bibinfo{journal}{Phys.\
  Rev.\ C} \textbf{\bibinfo{volume}{61}}, \bibinfo{pages}{014606}
  (\bibinfo{year}{1999}).

\bibitem[{\citenamefont{Piantelli et~al.}(2002)\citenamefont{Piantelli, Bidini,
  Poggi, Bini, Casini, Maurenzig, Olmi, Pasquali, Stefanini, and
  Taccetti}}]{Piantelli02}
\bibinfo{author}{\bibfnamefont{S.}~\bibnamefont{Piantelli}},
  \bibinfo{author}{\bibfnamefont{L.}~\bibnamefont{Bidini}},
  \bibinfo{author}{\bibfnamefont{G.}~\bibnamefont{Poggi}},
  \bibinfo{author}{\bibfnamefont{M.}~\bibnamefont{Bini}},
  \bibinfo{author}{\bibfnamefont{G.}~\bibnamefont{Casini}},
  \bibinfo{author}{\bibfnamefont{P.~R.} \bibnamefont{Maurenzig}},
  \bibinfo{author}{\bibfnamefont{A.}~\bibnamefont{Olmi}},
  \bibinfo{author}{\bibfnamefont{G.}~\bibnamefont{Pasquali}},
  \bibinfo{author}{\bibfnamefont{A.~A.} \bibnamefont{Stefanini}},
  \bibnamefont{and} \bibinfo{author}{\bibfnamefont{N.}~\bibnamefont{Taccetti}},
  \bibinfo{journal}{Phys.\ Rev.\ Lett.} \textbf{\bibinfo{volume}{88}},
  \bibinfo{pages}{052701} (\bibinfo{year}{2002}).

\bibitem[{\citenamefont{{Milazzo \textit{et al.}}}(2002)}]{Milazzo02}
\bibinfo{author}{\bibfnamefont{P.~M.} \bibnamefont{{Milazzo \textit{et al.}}}},
  \bibinfo{journal}{Nucl.\ Phys.} \textbf{\bibinfo{volume}{A703}},
  \bibinfo{pages}{466} (\bibinfo{year}{2002}).

\bibitem[{\citenamefont{{Mangiarotti \textit{et al.}}}(2004)}]{Mangiarotti04}
\bibinfo{author}{\bibfnamefont{A.}~\bibnamefont{{Mangiarotti \textit{et
  al.}}}}, \bibinfo{journal}{Phys.\ Rev.\ Lett.} \textbf{\bibinfo{volume}{93}},
  \bibinfo{pages}{232701} (\bibinfo{year}{2004}).

\bibitem[{\citenamefont{{Stefanini \textit{et al.}}}(1995)}]{Stefanini95}
\bibinfo{author}{\bibfnamefont{A.~A.} \bibnamefont{{Stefanini \textit{et
  al.}}}}, \bibinfo{journal}{Z.\ Phys.\ A} \textbf{\bibinfo{volume}{351}},
  \bibinfo{pages}{167} (\bibinfo{year}{1995}).

\bibitem[{\citenamefont{{Bowman \textit{et al.}}}(1993)}]{Bowman93}
\bibinfo{author}{\bibfnamefont{D.~R.} \bibnamefont{{Bowman \textit{et al.}}}},
  \bibinfo{journal}{Phys.\ Rev.\ Lett.} \textbf{\bibinfo{volume}{70}},
  \bibinfo{pages}{3534} (\bibinfo{year}{1993}).

\bibitem[{\citenamefont{{Montoya \textit{et al.}}}(1994)}]{Montoya94}
\bibinfo{author}{\bibfnamefont{C.~P.} \bibnamefont{{Montoya \textit{et al.}}}},
  \bibinfo{journal}{Phys.\ Rev.\ Lett.} \textbf{\bibinfo{volume}{73}},
  \bibinfo{pages}{3070} (\bibinfo{year}{1994}).

\bibitem[{\citenamefont{{T\~oke \textit{et al.}}}(1996)}]{Toke96}
\bibinfo{author}{\bibfnamefont{J.}~\bibnamefont{{T\~oke \textit{et al.}}}},
  \bibinfo{journal}{Phys.\ Rev.\ Lett.} \textbf{\bibinfo{volume}{77}},
  \bibinfo{pages}{3514} (\bibinfo{year}{1996}).

\bibitem[{\citenamefont{{Dempsey \textit{et al.}}}(1996)}]{Dempsey96}
\bibinfo{author}{\bibfnamefont{J.~F.} \bibnamefont{{Dempsey \textit{et al.}}}},
  \bibinfo{journal}{Phys.\ Rev.\ C} \textbf{\bibinfo{volume}{54}},
  \bibinfo{pages}{1710} (\bibinfo{year}{1996}).

\bibitem[{\citenamefont{Baran et~al.}(2004)\citenamefont{Baran, Colonna, and
  {Di Toro}}}]{Baran04}
\bibinfo{author}{\bibfnamefont{V.}~\bibnamefont{Baran}},
  \bibinfo{author}{\bibfnamefont{M.}~\bibnamefont{Colonna}}, \bibnamefont{and}
  \bibinfo{author}{\bibfnamefont{M.}~\bibnamefont{{Di Toro}}},
  \bibinfo{journal}{Nucl.\ Phys.} \textbf{\bibinfo{volume}{A730}},
  \bibinfo{pages}{329} (\bibinfo{year}{2004}).

\bibitem[{\citenamefont{Botvina et~al.}(1999)\citenamefont{Botvina, Bruno,
  D'Agostino, and Gross}}]{Botvina99}
\bibinfo{author}{\bibfnamefont{A.~S.} \bibnamefont{Botvina}},
  \bibinfo{author}{\bibfnamefont{M.}~\bibnamefont{Bruno}},
  \bibinfo{author}{\bibfnamefont{M.}~\bibnamefont{D'Agostino}},
  \bibnamefont{and} \bibinfo{author}{\bibfnamefont{D.~H.~E.}
  \bibnamefont{Gross}}, \bibinfo{journal}{Phys.\ Rev.\ C}
  \textbf{\bibinfo{volume}{59}}, \bibinfo{pages}{3444} (\bibinfo{year}{1999}).

\bibitem[{\citenamefont{Botvina and Mishustin}(2001)}]{Botvina01}
\bibinfo{author}{\bibfnamefont{A.~S.} \bibnamefont{Botvina}} \bibnamefont{and}
  \bibinfo{author}{\bibfnamefont{I.~N.} \bibnamefont{Mishustin}},
  \bibinfo{journal}{Phys.\ Rev.\ C} \textbf{\bibinfo{volume}{63}},
  \bibinfo{pages}{61601R} (\bibinfo{year}{2001}).

\bibitem[{\citenamefont{{Casini \textit{et al.}}}(1993)}]{Casini93}
\bibinfo{author}{\bibfnamefont{G.}~\bibnamefont{{Casini \textit{et al.}}}},
  \bibinfo{journal}{Phys.\ Rev.\ Lett.} \textbf{\bibinfo{volume}{71}},
  \bibinfo{pages}{2567} (\bibinfo{year}{1993}).

\bibitem[{\citenamefont{{Bocage \textit{et al.}}}(2000)}]{Bocage00}
\bibinfo{author}{\bibfnamefont{F.}~\bibnamefont{{Bocage \textit{et al.}}}}
  (\bibinfo{collaboration}{INDRA collaboration}), \bibinfo{journal}{Nucl.\
  Phys.} \textbf{\bibinfo{volume}{A676}}, \bibinfo{pages}{391}
  (\bibinfo{year}{2000}).

\bibitem[{\citenamefont{{De Filippo \textit{et al.}}}(2005)}]{Defilippo05neck}
\bibinfo{author}{\bibfnamefont{E.}~\bibnamefont{{De Filippo \textit{et al.}}}},
  \bibinfo{journal}{Phys.\ Rev.\ C} \textbf{\bibinfo{volume}{71}},
  \bibinfo{pages}{044602} (\bibinfo{year}{2005}).

\bibitem[{\citenamefont{{Di Toro} et~al.}(2006)\citenamefont{{Di Toro}, Olmi,
  and Roy}}]{WCI06}
\bibinfo{author}{\bibfnamefont{M.}~\bibnamefont{{Di Toro}}},
  \bibinfo{author}{\bibfnamefont{A.}~\bibnamefont{Olmi}}, \bibnamefont{and}
  \bibinfo{author}{\bibfnamefont{R.}~\bibnamefont{Roy}},
  \bibinfo{journal}{Eur.\ Phys.\ J.\ A}  (\bibinfo{year}{2006}),
  \bibinfo{note}{(to be published)}.

\bibitem[{\citenamefont{{Hudan \textit{et al.}}}(2005)}]{Hudan05}
\bibinfo{author}{\bibfnamefont{S.}~\bibnamefont{{Hudan \textit{et al.}}}},
  \bibinfo{journal}{Phys.\ Rev.\ C} \textbf{\bibinfo{volume}{71}},
  \bibinfo{pages}{054604} (\bibinfo{year}{2005}).

\bibitem[{\citenamefont{Durand}(1998)}]{Durand98}
\bibinfo{author}{\bibfnamefont{D.}~\bibnamefont{Durand}},
  \bibinfo{journal}{Nucl.\ Phys.} \textbf{\bibinfo{volume}{A630}},
  \bibinfo{pages}{52c} (\bibinfo{year}{1998}).

\bibitem[{\citenamefont{{Bini \textit{et al.}}}(2003)}]{Bini03}
\bibinfo{author}{\bibfnamefont{M.}~\bibnamefont{{Bini \textit{et al.}}}},
  \bibinfo{journal}{Nucl.\ Instrum.\ Methods Phys.\ Res.\ A}
  \textbf{\bibinfo{volume}{515}}, \bibinfo{pages}{497} (\bibinfo{year}{2003}).

\bibitem[{\citenamefont{Piantelli}(2001)}]{Piantelli01}
\bibinfo{author}{\bibfnamefont{S.}~\bibnamefont{Piantelli}}, Ph.D. thesis,
  \bibinfo{school}{Florence University} (\bibinfo{year}{2001}).

\bibitem[{\citenamefont{Mangiarotti}(2003)}]{Mangiarotti03}
\bibinfo{author}{\bibfnamefont{A.}~\bibnamefont{Mangiarotti}}, Ph.D. thesis,
  \bibinfo{school}{Florence University} (\bibinfo{year}{2003}).

\bibitem[{\citenamefont{Bass}(1980)}]{Bass80}
\bibinfo{author}{\bibfnamefont{R.}~\bibnamefont{Bass}},
  \emph{\bibinfo{title}{Nuclear Reactions with Heavy Ions}}, Texts and
  Monographs in Physics (\bibinfo{publisher}{Springer Verlag},
  \bibinfo{year}{1980}).

\bibitem[{\citenamefont{{Charity \textit{et al.}}}(1991)}]{Charity91}
\bibinfo{author}{\bibfnamefont{R.~J.} \bibnamefont{{Charity \textit{et al.}}}},
  \bibinfo{journal}{Z.\ Phys.\ A} \textbf{\bibinfo{volume}{341}},
  \bibinfo{pages}{53} (\bibinfo{year}{1991}).

\bibitem[{\citenamefont{Casini et~al.}(1989)\citenamefont{Casini, Maurenzig,
  Olmi, and Stefanini}}]{Casini89}
\bibinfo{author}{\bibfnamefont{G.}~\bibnamefont{Casini}},
  \bibinfo{author}{\bibfnamefont{P.~R.} \bibnamefont{Maurenzig}},
  \bibinfo{author}{\bibfnamefont{A.}~\bibnamefont{Olmi}}, \bibnamefont{and}
  \bibinfo{author}{\bibfnamefont{A.~A.} \bibnamefont{Stefanini}},
  \bibinfo{journal}{Nucl.\ Instrum.\ Methods Phys.\ Res.\ A}
  \textbf{\bibinfo{volume}{277}}, \bibinfo{pages}{445} (\bibinfo{year}{1989}).

\bibitem[{\citenamefont{{Imm{\`e} \textit{et al.}}}(March 1993)}]{Racitihodo}
\bibinfo{author}{\bibfnamefont{G.}~\bibnamefont{{Imm{\`e} \textit{et al.}}}},
  in \emph{\bibinfo{booktitle}{Proc. Workshop on Detector and Instrumentation,
  Acireale (Catania)}} (\bibinfo{year}{March 1993}).

\bibitem[{\citenamefont{Sfienti et~al.}(2004)\citenamefont{Sfienti, Baran, {De
  Napoli}, Imm{\'e}, Raciti, Rapisarda, Rascun{\'a}, and Spezzi}}]{Sfienti04}
\bibinfo{author}{\bibfnamefont{C.}~\bibnamefont{Sfienti}},
  \bibinfo{author}{\bibfnamefont{V.}~\bibnamefont{Baran}},
  \bibinfo{author}{\bibfnamefont{M.}~\bibnamefont{{De Napoli}}},
  \bibinfo{author}{\bibfnamefont{G.}~\bibnamefont{Imm{\'e}}},
  \bibinfo{author}{\bibfnamefont{G.}~\bibnamefont{Raciti}},
  \bibinfo{author}{\bibfnamefont{E.}~\bibnamefont{Rapisarda}},
  \bibinfo{author}{\bibfnamefont{S.}~\bibnamefont{Rascun{\'a}}},
  \bibnamefont{and} \bibinfo{author}{\bibfnamefont{L.}~\bibnamefont{Spezzi}},
  \bibinfo{journal}{Nucl.\ Phys.} \textbf{\bibinfo{volume}{A734}},
  \bibinfo{pages}{528} (\bibinfo{year}{2004}).

\bibitem[{\citenamefont{Schr{\"o}der and Huizenga}(1977)}]{Schroeder77}
\bibinfo{author}{\bibfnamefont{W.~U.} \bibnamefont{Schr{\"o}der}}
  \bibnamefont{and} \bibinfo{author}{\bibfnamefont{J.~R.}
  \bibnamefont{Huizenga}}, \bibinfo{journal}{Ann. Rev. Nucl. Sci.}
  \textbf{\bibinfo{volume}{27}}, \bibinfo{pages}{465} (\bibinfo{year}{1977}).

\bibitem[{\citenamefont{{D'Agostino \textit{et al.}}}(1999)}]{Dagostino99}
\bibinfo{author}{\bibfnamefont{M.}~\bibnamefont{{D'Agostino \textit{et al.}}}},
  \bibinfo{journal}{Nucl.\ Phys.} \textbf{\bibinfo{volume}{A650}},
  \bibinfo{pages}{329} (\bibinfo{year}{1999}).

\bibitem[{\citenamefont{{Mangiarotti \textit{et al.}}}()}]{Mangiarotti05}
\bibinfo{author}{\bibfnamefont{A.}~\bibnamefont{{Mangiarotti \textit{et
  al.}}}}, \bibinfo{note}{to be published}.

\bibitem[{\citenamefont{D\mbox{\o}ssing}(1981)}]{Dossing81}
\bibinfo{author}{\bibfnamefont{T.}~\bibnamefont{D\mbox{\o}ssing}},
  \bibinfo{journal}{Nucl.\ Phys.} \textbf{\bibinfo{volume}{A357}},
  \bibinfo{pages}{488} (\bibinfo{year}{1981}).

\bibitem[{\citenamefont{Moretto et~al.}(1981)\citenamefont{Moretto, Blau, and
  Pacheco}}]{Moretto81}
\bibinfo{author}{\bibfnamefont{L.~G.} \bibnamefont{Moretto}},
  \bibinfo{author}{\bibfnamefont{S.~K.} \bibnamefont{Blau}}, \bibnamefont{and}
  \bibinfo{author}{\bibfnamefont{A.~J.} \bibnamefont{Pacheco}},
  \bibinfo{journal}{Nucl.\ Phys.} \textbf{\bibinfo{volume}{A364}},
  \bibinfo{pages}{125} (\bibinfo{year}{1981}).

\bibitem[{\citenamefont{{Charity et al.}}(1988)}]{Charity88b}
\bibinfo{author}{\bibfnamefont{R.~J.} \bibnamefont{{Charity et al.}}},
  \bibinfo{journal}{Nucl.\ Phys.} \textbf{\bibinfo{volume}{A483}},
  \bibinfo{pages}{371} (\bibinfo{year}{1988}).

\bibitem[{\citenamefont{Bondorf et~al.}(1995)\citenamefont{Bondorf, Botvina,
  Iljinov, Mishustin, and Sneppen}}]{Bondorf95}
\bibinfo{author}{\bibfnamefont{J.~P.} \bibnamefont{Bondorf}},
  \bibinfo{author}{\bibfnamefont{A.~S.} \bibnamefont{Botvina}},
  \bibinfo{author}{\bibfnamefont{A.~S.} \bibnamefont{Iljinov}},
  \bibinfo{author}{\bibfnamefont{I.~N.} \bibnamefont{Mishustin}},
  \bibnamefont{and} \bibinfo{author}{\bibfnamefont{K.}~\bibnamefont{Sneppen}},
  \bibinfo{journal}{Phys. Rep.} \textbf{\bibinfo{volume}{257}},
  \bibinfo{pages}{133} (\bibinfo{year}{1995}).

\bibitem[{\citenamefont{{Larochelle \textit{et al.}}}(2000)}]{Larochelle00}
\bibinfo{author}{\bibfnamefont{Y.}~\bibnamefont{{Larochelle \textit{et al.}}}},
  \bibinfo{journal}{Phys.\ Rev.\ C} \textbf{\bibinfo{volume}{62}},
  \bibinfo{pages}{51602R} (\bibinfo{year}{2000}).

\bibitem[{\citenamefont{Moretto et~al.}(1993)\citenamefont{Moretto, Delis, and
  Wozniak}}]{Moretto93}
\bibinfo{author}{\bibfnamefont{L.~G.} \bibnamefont{Moretto}},
  \bibinfo{author}{\bibfnamefont{D.~N.} \bibnamefont{Delis}}, \bibnamefont{and}
  \bibinfo{author}{\bibfnamefont{G.~J.} \bibnamefont{Wozniak}},
  \bibinfo{journal}{Phys.\ Rev.\ Lett.} \textbf{\bibinfo{volume}{71}},
  \bibinfo{pages}{3935} (\bibinfo{year}{1993}).

\bibitem[{\citenamefont{Botvina and Gross}(1995)}]{Botvina95}
\bibinfo{author}{\bibfnamefont{A.~S.} \bibnamefont{Botvina}} \bibnamefont{and}
  \bibinfo{author}{\bibfnamefont{D.~H.~E.} \bibnamefont{Gross}},
  \bibinfo{journal}{Phys.\ Lett.} \textbf{\bibinfo{volume}{B344}},
  \bibinfo{pages}{6} (\bibinfo{year}{1995}).

\bibitem[{\citenamefont{{Elliott \textit{et al.}}}(2000)}]{Elliott00}
\bibinfo{author}{\bibfnamefont{J.~B.} \bibnamefont{{Elliott \textit{et al.}}}},
  \bibinfo{journal}{Phys.\ Rev.\ Lett.} \textbf{\bibinfo{volume}{85}},
  \bibinfo{pages}{1194} (\bibinfo{year}{2000}).

\bibitem[{\citenamefont{{Beaulieu \textit{et al.}}}(2001)}]{Beaulieu01}
\bibinfo{author}{\bibfnamefont{L.}~\bibnamefont{{Beaulieu \textit{et al.}}}},
  \bibinfo{journal}{Phys.\ Rev.\ C} \textbf{\bibinfo{volume}{63}},
  \bibinfo{pages}{031302} (\bibinfo{year}{2001}).

\bibitem[{\citenamefont{{Djerroud \textit{et al.}}}(2001)}]{Djerroud01}
\bibinfo{author}{\bibfnamefont{B.}~\bibnamefont{{Djerroud \textit{et al.}}}},
  \bibinfo{journal}{Phys.\ Rev.\ C} \textbf{\bibinfo{volume}{64}},
  \bibinfo{pages}{034603} (\bibinfo{year}{2001}).

\bibitem[{\citenamefont{{Yanez \textit{et al.}}}(2003)}]{Yanez03}
\bibinfo{author}{\bibfnamefont{R.}~\bibnamefont{{Yanez \textit{et al.}}}},
  \bibinfo{journal}{Phys.\ Rev.\ C} \textbf{\bibinfo{volume}{68}},
  \bibinfo{pages}{011602} (\bibinfo{year}{2003}).

\bibitem[{\citenamefont{{M\'etivier \textit{et al.}}}(2000)}]{Metivier00}
\bibinfo{author}{\bibfnamefont{V.}~\bibnamefont{{M\'etivier \textit{et al.}}}}
  (\bibinfo{collaboration}{INDRA collaboration}), \bibinfo{journal}{Nucl.\
  Phys.} \textbf{\bibinfo{volume}{A672}}, \bibinfo{pages}{357}
  (\bibinfo{year}{2000}).

\bibitem[{\citenamefont{{\L}ukasik and Majka}(1993)}]{Lukasik93}
\bibinfo{author}{\bibfnamefont{J.}~\bibnamefont{{\L}ukasik}} \bibnamefont{and}
  \bibinfo{author}{\bibfnamefont{Z.}~\bibnamefont{Majka}},
  \bibinfo{journal}{Acta Phys. Pol.} \textbf{\bibinfo{volume}{B24}},
  \bibinfo{pages}{1959} (\bibinfo{year}{1993}).

\bibitem[{\citenamefont{Schr{\"o}der et~al.}(1978)\citenamefont{Schr{\"o}der,
  Birkelund, Huizenga, Wolf, and {Viola Jr.}}}]{Schroeder78}
\bibinfo{author}{\bibfnamefont{W.~U.} \bibnamefont{Schr{\"o}der}},
  \bibinfo{author}{\bibfnamefont{J.~R.} \bibnamefont{Birkelund}},
  \bibinfo{author}{\bibfnamefont{J.~R.} \bibnamefont{Huizenga}},
  \bibinfo{author}{\bibfnamefont{K.~L.} \bibnamefont{Wolf}}, \bibnamefont{and}
  \bibinfo{author}{\bibfnamefont{V.~E.} \bibnamefont{{Viola Jr.}}},
  \bibinfo{journal}{Phys. Rep.} \textbf{\bibinfo{volume}{45}},
  \bibinfo{pages}{301} (\bibinfo{year}{1978}).

\bibitem[{\citenamefont{Wilczynski}(1973)}]{Wilczynski73}
\bibinfo{author}{\bibfnamefont{J.}~\bibnamefont{Wilczynski}},
  \bibinfo{journal}{Phys.\ Lett.} \textbf{\bibinfo{volume}{B47}},
  \bibinfo{pages}{484} (\bibinfo{year}{1973}).

\end{thebibliography}

\end{document}